\documentclass[trackchanges, twocolumn]{aastex701}

\usepackage{graphicx} 
\usepackage{subcaption} 
\usepackage{amsmath}
\usepackage{booktabs}
\usepackage{url}
\usepackage{natbib}
\usepackage{multirow} 
\begin{document}

\title{The annular gap model under a rotating dipole field approximation: simulating gamma-ray light curve}

\author[orcid=0009-0001-5676-3926]{Jie Tian}
\affiliation{School of Mathematical Science, Guizhou Normal University, Guiyang 550001, China}
\affiliation{Guizhou Provincial Key Laboratory of Radio Astronomy and Data Processing, Guizhou Normal University, Guiyang 550001, China}
\email{jietian2024@163.com}

\author[orcid=0009-0006-3224-4319]{Xin Xu}
\affiliation{School of Mathematical Science, Guizhou Normal University, Guiyang 550001, China}
\affiliation{Guizhou Provincial Key Laboratory of Radio Astronomy and Data Processing, Guizhou Normal University, Guiyang 550001, China}
\email{xuxin_nl@163.com}

\author[orcid=0000-0001-9389-5197]{Qijun Zhi}
\affiliation{School of Science, Guizhou University, Guiyang 550001, China}
\affiliation{School of Physics and Electronic Science, Guizhou Normal University, Guiyang, 550001, China}
\affiliation{Guizhou Provincial Key Laboratory of Radio Astronomy and Data Processing, Guizhou Normal University, Guiyang 550001, China}
\email{qjzhi@gznu.edu.cn}

\author[orcid=0000-0002-9815-5873]{Jiguang Lu}
\affiliation{Guizhou Radio Astronomical Observatory, Guiyang 550025, China}
\affiliation{National Astronomical Observatories, Chinese Academy of Sciences, Beijing 100012, China}
\email{lujig@nao.cas.cn}

\author[orcid=0000-0002-2060-5539]{Shijun Dang}
\affiliation{Guizhou Provincial Key Laboratory of Radio Astronomy and Data Processing, Guizhou Normal University, Guiyang 550001, China}
\affiliation{School of Physics and Electronic Science, Guizhou Normal University, Guiyang, 550001, China}
\email{dangsj@gznu.edu.cn}

\author{Ke Yang}
\affiliation{Guizhou Provincial Key Laboratory of Radio Astronomy and Data Processing, Guizhou Normal University, Guiyang 550001, China}
\affiliation{School of Physics and Electronic Science, Guizhou Normal University, Guiyang, 550001, China}
\email{yangk1632617@163.com}

\author{Xiao Wei}
\affiliation{School of Science, Guizhou University, Guiyang 550001, China}
\email{xiaowei2004@163.com}

\author{Guojun Qiao}
\affiliation{School of Physics, Peking University, Beijing 100871, China}
\email{gjn@pku.edu.cn}

\correspondingauthor{Qijun Zhi}
\email{qjzhi@gznu.edu.cn}

\correspondingauthor{Jiguang Lu}
\email{lujig@nao.cas.cn}

\correspondingauthor{Shijun Dang}
\email{dangsj@gznu.edu.cn}

\begin{abstract}
A more realistic description of the magnetosphere is crucial for understanding the radiation emitted by pulsars.
In this paper, we revisit the annular gap model by employing a rotating dipole field, which is more realistic than the static dipole field,  as an approximation of the magnetic structure of the pulsar magnetosphere.
Compared with the static dipole field approximation, the open field-line region, including both the core and annular gaps, is significantly enlarged, and the two regions become asymmetric with respect to the fiducial plane. 
We apply this model to three young gamma-ray pulsars with distinct light-curve morphologies, PSRs J0631$+$1036 (single peak), J1709$-$4429 (double peaks), and J1048$-$5832 (three peaks). 
Using viewing geometries constrained by radio polarization measurements, the annular gap model within the rotating dipole field successfully reproduces the main morphological features of their gamma-ray light curves above 0.1 GeV. 
Our model provides a framework for interpreting pulsar high-energy emission, which can be used to analyze the emission properties of high-energy pulsars.
\end{abstract}

\keywords{\uat{Pulsars}{1306} --- \uat{Magnetic fields}{994} --- \uat{High Energy astrophysics}{739}}

\section{Introduction} \label{sec:1}
Pulsars (PSRs) are highly magnetized, rapidly rotating neutron stars, with approximately four thousand discovered to date \citep{2005AJ....129.1993M, 2024ApJ...960...79ZhiQJ, 2025ApJ...982..117XuX, 2025RAA....25a4001HanJL}.
High-energy pulsars, typically young or millisecond pulsars, form a subclass that radiates electromagnetic waves in high-energy bands, and some also emit at other wavelengths. In the Third Fermi gamma-ray pulsar catalog\citep[][hereafter 3PC]{2023ApJ...958..191S}, the light curves exhibit a variety of pulse profile types: single-peak, double-peak, and multi-peak profiles, with corresponding proportions of 26\%, 53\%, and 21\%, respectively. Among the 120 Fermi pulsars with double-peak profiles, the distribution of peak separation ($\Delta$) is non-uniform, ranging from 0.082 to 0.73, with about 55\% having $\Delta \geq 0.4$. Meanwhile, the observed gamma-ray spectra are generally described by an exponentially cutoff power law, with cutoff energies typically between 1 and 5 GeV. These observed features reflect the intrinsic characteristics of high-energy pulsar radiation and contribute to the study of their radiation mechanisms and geometry.

The structure of the magnetic field in the pulsar magnetosphere plays a crucial role in exploring pulsar radiation mechanisms and potential emission sites, serving as a key to unlocking many of the mysteries inherent in these celestial objects. With advances in numerical techniques and increased computational power, the pulsar magnetosphere has been extensively simulated and studied\citep[e.g.][] {2006ApJ...648L..51S,2013MNRAS.435L...1T,2016MNRAS.455.3779P,2018PhRvD..98b3010C,2012ApJ...749....2K,2015ApJ...801L..19P,2014ApJ...785L..33P,2018ApJ...855...94P}. It is generally considered that a rotating(vacuum retarded) dipole field \citep{1955AnAp...18....1D,2000ApJ...537..964C} and a force-free field \citep[e.g.][]{2006MNRAS.368.1055T,2016MNRAS.455.3779P} represent two extreme theoretical cases for the pulsar magnetosphere, with real magnetospheres expected to lie between them, such as in resistive magnetosphere models \citep[e.g.][]{2012ApJ...746...60L}. 

To explain the observed multi-band electromagnetic emissions from pulsars, a particle acceleration region is required. Therefore, various theoretical models have been proposed, which can be divided into two categories based on the location of this region. In the first category, the region that accelerates primary particles and produces high-energy emission lies inside the light cylinder, examples include the polar cap model \citep{1975ApJ...196...51R,1982ApJ...252..337D}, the slot gap model \citep{1983ApJ...266..215A}, the two-pole caustic model \citep{2003ApJ...598.1201D}, the outer gap model \citep{1986ApJ...300..500C,1986ApJ...300..522C}, the pair-starved polar cap model \citep{2005ApJ...622..531H} and the angular model\citep{2004ApJ...606L..49Q,2010MNRAS.406.2671D}. In the second category, the acceleration region and the resulting high-energy emission are situated near the equatorial current sheet outside the light cylinder. This group includes the striped-wind models \citep{1983ZhETF..85..401B,1990ApJ...349..538C}, the force-free inside/dissipative outside model \citep[e.g.][]{2014ApJ...793...97K,2017ApJ...842...80K}, and kinetic or particle-in-cell models \citep[e.g.][]{2018ApJ...857...44K,2018ApJ...855...94P}. These models have been widely used by researchers to simulate light curves and spectra of high-energy pulsars \citep[e.g.][]{2009ApJ...707..800V,2012ApJ...744...34V,2014ApJS..213....6J,2015A&A...575A...3P,2016A&A...588A.137P,2010ApJ...715.1282B,2021A&A...647A.101B,2021A&A...654A.106P,2022ApJ...925..184B,2019ApJ...874..166C,2017ApJ...842...80K,2018ApJ...855...94P,2016MNRAS.457.2401C}.

The annular gap model, by assuming a static dipole field approximation for the magnetic field of the pulsar magnetosphere, was initially proposed by \cite{2004ApJ...606L..49Q} and then developed by \citet{2010MNRAS.406.2671D}. In this model, due to the limitation of the speed of light, charged particles flow outward along the magnetic field lines. To accommodate the changes in the flow of charged particles, a gap forms in the magnetosphere. Within this gap, the charge density deviates from the Goldreich–Julian\citep[GJ,][]{1969ApJ...157..869G} charge density, and this deviation induces an accelerating electric field, which can accelerate charged particles, leading to the observed electromagnetic radiation through complex physical mechanisms. 
This model simplifies the treatment of plasma physical processes and the pulsar magnetosphere structure, and is a semi-analytical model. It has been applied to study several pulsars and has reproduced their light curves and/or spectra\citep[e.g.][]{2011ApJ...731....2D,2012ApJ...748...84D,2013ApJ...763...29D,2015ApJ...801..131D}. 
In fact, for the pulsar magnetosphere, the rotating dipole field, rather than the static dipole field, is more  realistic and has been adopted in other models\citep[e.g. ][]{2000ApJ...537..964C, 2018MNRAS.475.2185C, 2019MNRAS.488.4288C}.
In this study, we aim to refine this geometric framework by adopting the realistic rotating dipole field and using it to study the light curves of high-energy pulsars. This paper is structured as follows: In Section \ref{sec:2}, we describe the geometry of a rotating dipole field magnetosphere. In Section \ref{sec:3}, we introduce the physics and modeling of the annular gap model. We present the results of applying this model to the pulsars with distinct gamma-ray light-curve profiles in Section \ref{sec:4}. Finally, the discussions and conclusions are presented in Section \ref{sec:5}.

\section{Magnetospheric geometry} \label{sec:2}
\begin{figure}
    \centering
    \includegraphics[width=0.8\linewidth]{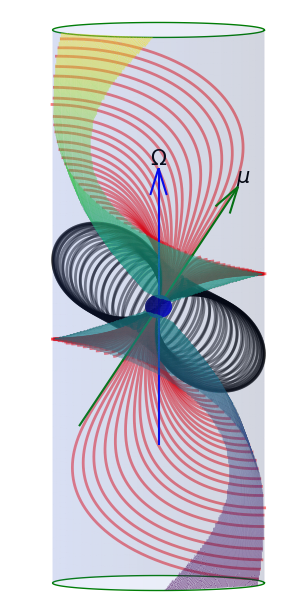}
    \caption{
    An example of a rotating dipole magnetosphere plotted using Crab pulsar parameters and $\alpha=30^\circ$.
    The red and black curves represent the critical field lines and the last open field lines, respectively. The colored surface and the grey cylindrical surface depict the null charge surface and the light cylinder, respectively.}
    \label{fig:1}
\end{figure}

We adopt a rotating dipole field as an approximation to the pulsar magnetospheric magnetic field. For a pulsar with angular velocity $\rm \Omega$ ($\Omega = \frac{\rm 2\pi}{\rm P}$, $\rm P$ is the pulsar spin period) and magnetic inclination angle $\alpha$, taking the $\hat{z}$-axis as the rotation axis, the local magnetic field can be written as 
\begin{equation}
    \textbf{B} = \hat{\textbf{r}} \left[\hat{\textbf{r}}\cdot\left(\frac{3\mu(t)}{r^3}+\frac{3\dot{\mu}(t)}{cr^2}+\frac{\ddot{\mu}(t)}{c^2r}\right)\right]-\left[\frac{\mu(t)}{r^3}+\frac{\dot{\mu}(t)}{cr^2}+\frac{\ddot{\mu}(t)}{c^2r}\right]
\end{equation}
\citep{2000ApJ...537..964C}, where the magnetic moment vector is given by $\mu(t) = \mu (\hat{\textbf{x}}\,\sin\alpha\, \text{cos}\Omega t\,+\, \hat{\textbf{y}}\,\sin\alpha\,\sin\Omega t\, +\, \hat{\textbf{z}}\,\text{cos}\alpha)$, $\hat{\textbf{r}}$ is the radial unit vector, and $c$ is light speed. The three components of a rotating dipole field in Cartesian and spherical coordinates can be seen in \citet{2004ApJ...614..869D}. The structure of the magnetosphere can still be described by a specific feature such as the last open field lines that tangent to the light cylinder of radius $R_{\rm LC} = c/\Omega$ and the critical field lines which cross the intersection of the null charge surface (\textbf{$\Omega \cdot B = 0 $}) and the light cylinder. Figure\,\ref{fig:1} shows an example of such a configuration for $\alpha = 30^\circ$. 

\begin{figure*}[htbp]
    \centering
    \begin{subfigure}[b]{0.245\textwidth} 
        \centering
        \includegraphics[width=\linewidth]{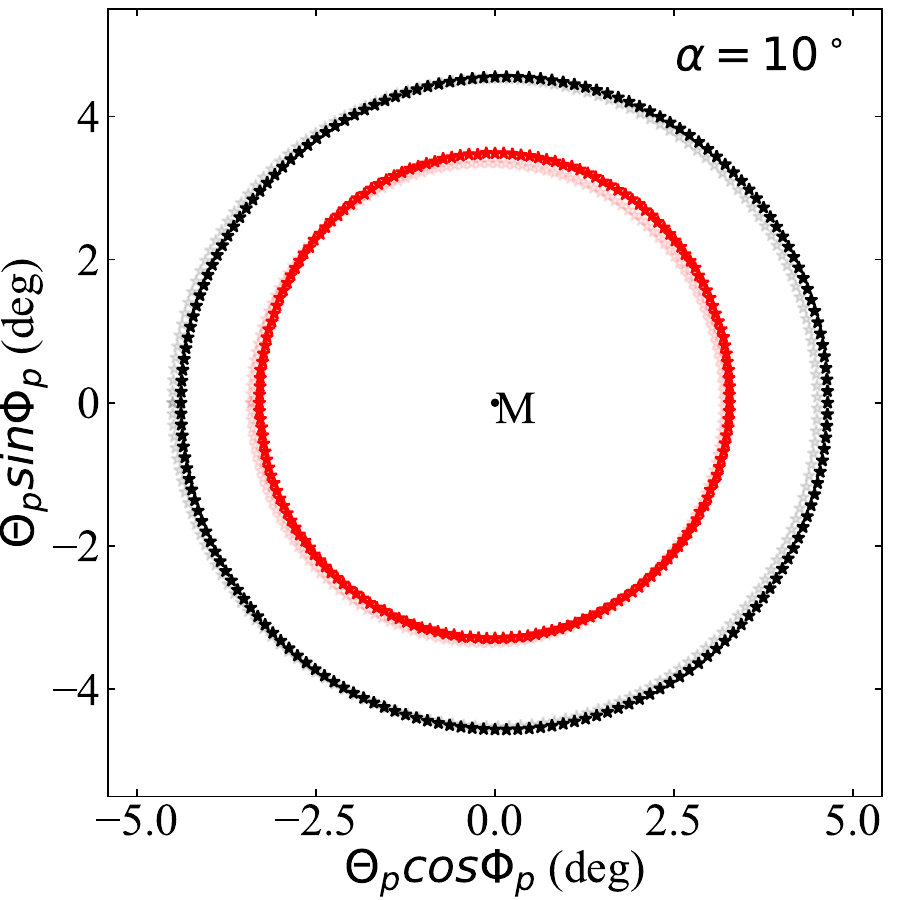} 
        \label{fig:sub1}
    \end{subfigure}
    \hfill
    \begin{subfigure}[b]{0.245\textwidth} 
        \centering
        \includegraphics[width=\linewidth]{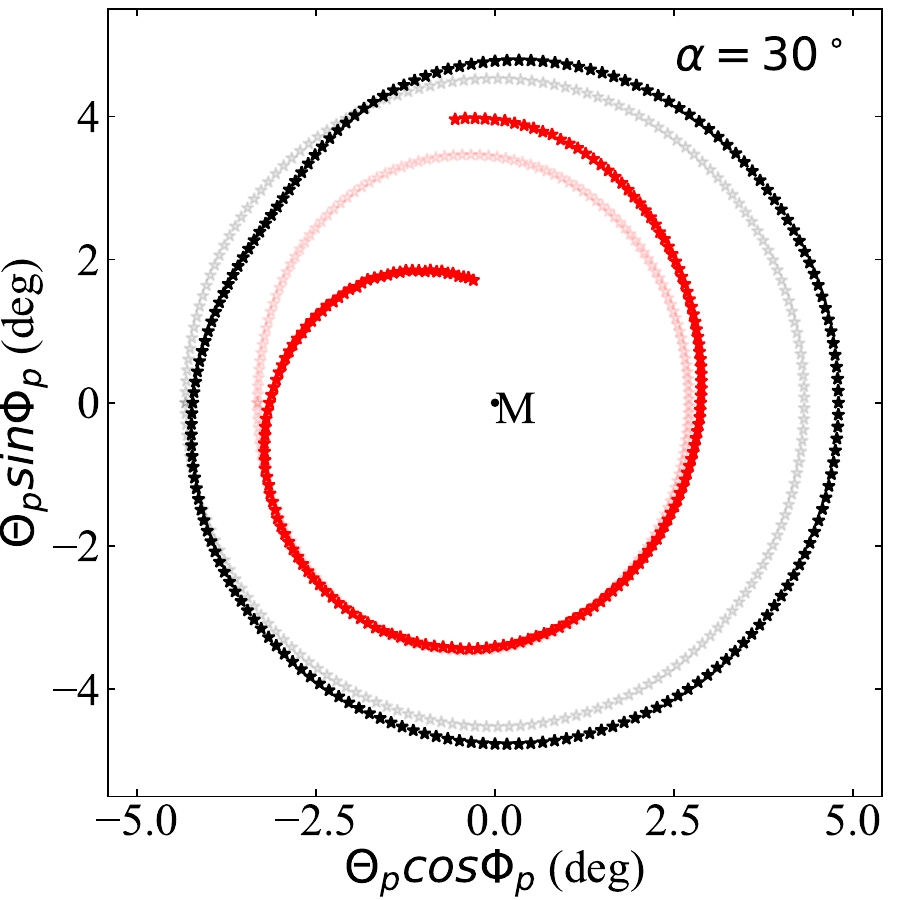} 
        \label{fig:sub1}
    \end{subfigure}
    \hfill 
    \begin{subfigure}[b]{0.245\textwidth} 
        \centering
        \includegraphics[width=\linewidth]{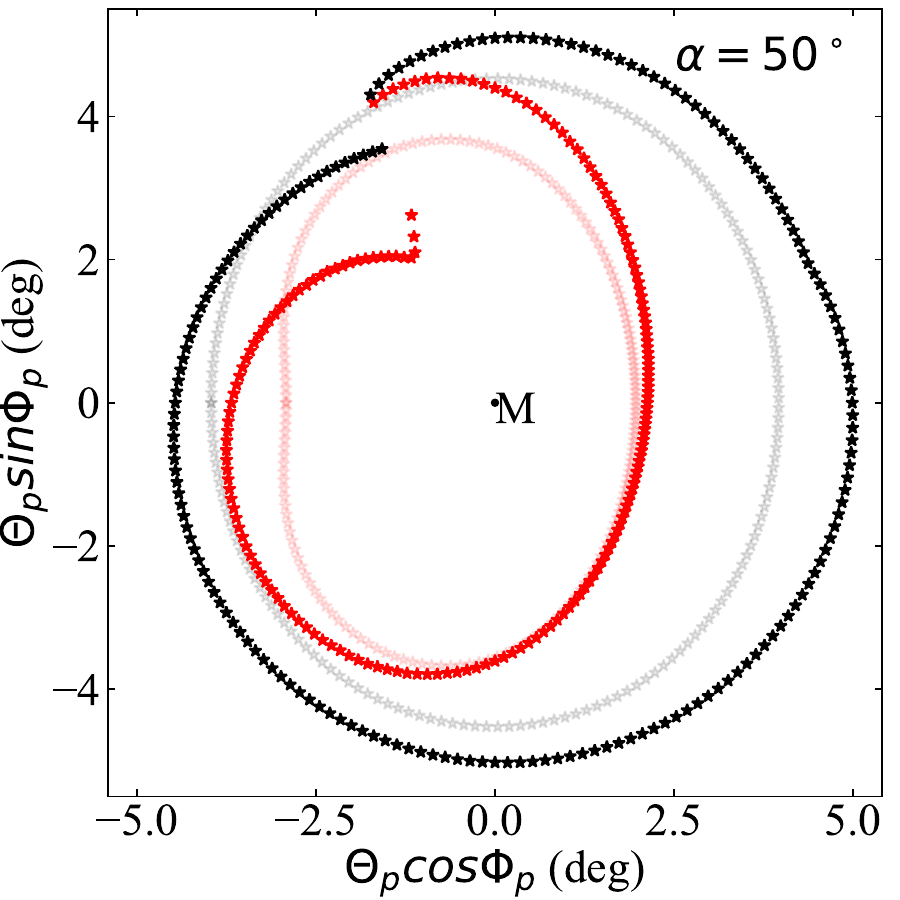} 
        \label{fig:sub1}
    \end{subfigure}
    \hfill
    \begin{subfigure}[b]{0.245\textwidth}
        \centering
        \includegraphics[width=\linewidth]{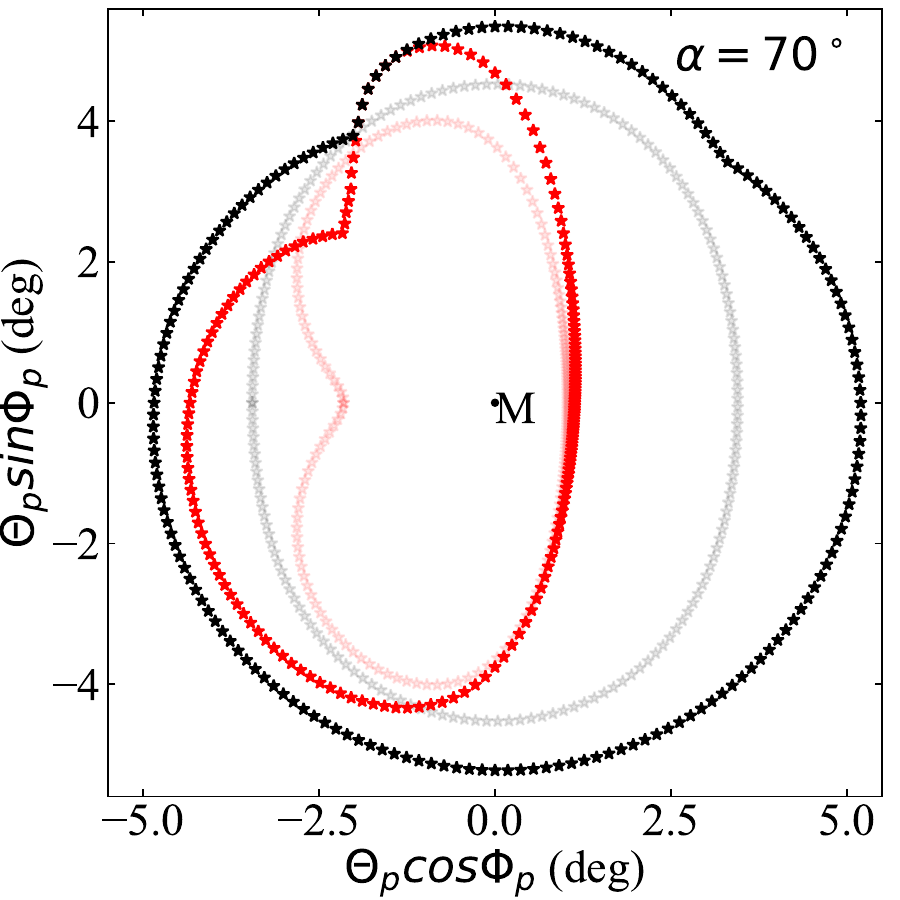}
        \label{fig:sub2}
    \end{subfigure}
    \caption{The diagram illustrates the footprints of the last open field lines (in black) and the critical field lines (in red), calculated using the Crab pulsar parameters and various $\alpha$, under both the rotating dipole approximation (dark shading) and the static dipole approximation (light shading). $\Theta_{\rm p}$ and $\Phi_{\rm p}$ are the magnetic colatitude and magnetic azimuth angle for footprints, respectively.}
    \label{fig:2}
\end{figure*}

The polar cap of a pulsar is defined by the footprints of the last open field lines on the stellar surface and can be divided into two regions by the footprints of the critical field lines.
For an aligned rotator whose magnetosphere is approximated as a static dipole field, the footprint trajectories of the critical and last open field lines are both circles. Their radii are given by $r_{\rm core} = \left(\frac{2}{3}\right) ^\frac{3}{4} R \left(\frac{\Omega R}{c}\right) ^\frac{1}{2}$ and $r_{\rm p} = R \left(\frac{\Omega R}{c}\right) ^\frac{1}{2}$, respectively\citep{1975ApJ...196...51R}, where $R$ is the neutron star radius. 
In Figure\,\ref{fig:2}, we show the footprints of the last open and critical field lines under both the rotating and static dipole field approximations, for different values of $\alpha$. 
It can be seen that the footprints of the last open and critical field lines in the rotating dipole case differ from those in the static dipole case, particularly at large $\alpha$.

\begin{figure*}[htbp]
    \centering
    \begin{subfigure}[b]{0.245\textwidth} 
        \centering
        \includegraphics[width=\linewidth]{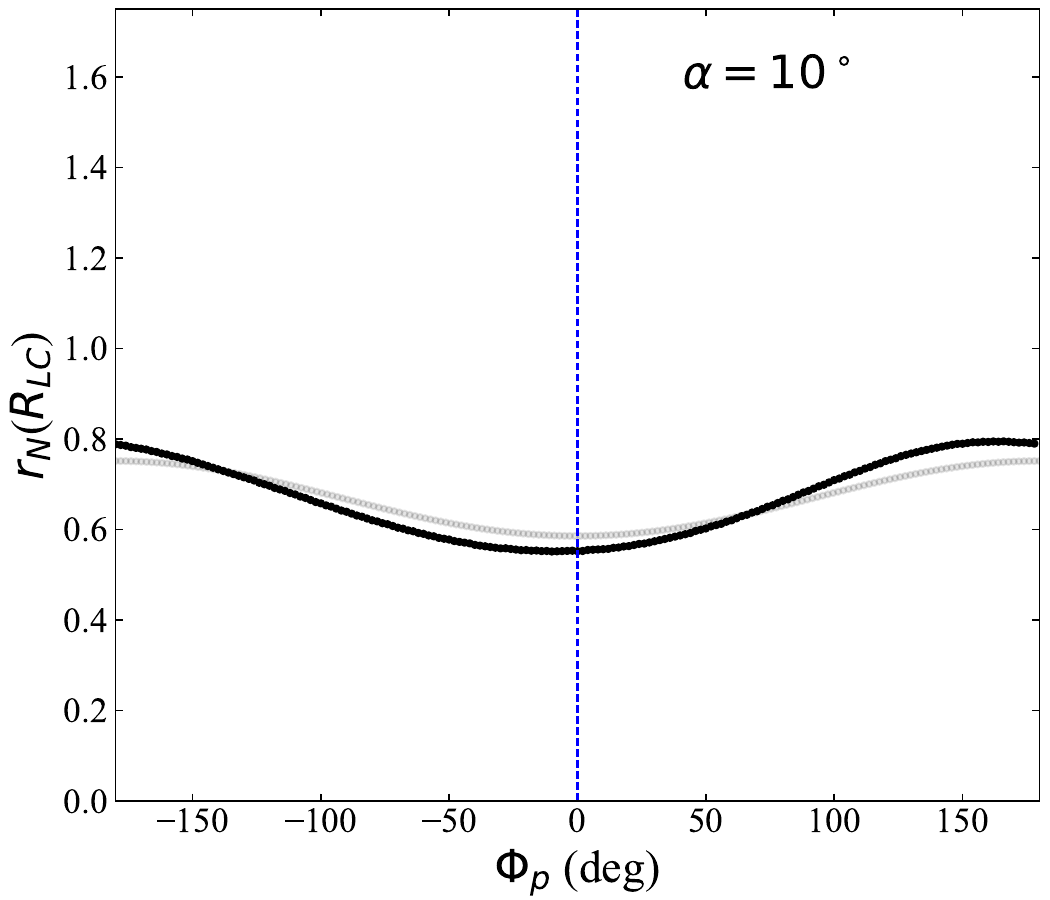} 
        \label{fig:sub1}
    \end{subfigure}
    \hfill
    \begin{subfigure}[b]{0.245\textwidth} 
        \centering
        \includegraphics[width=\linewidth]{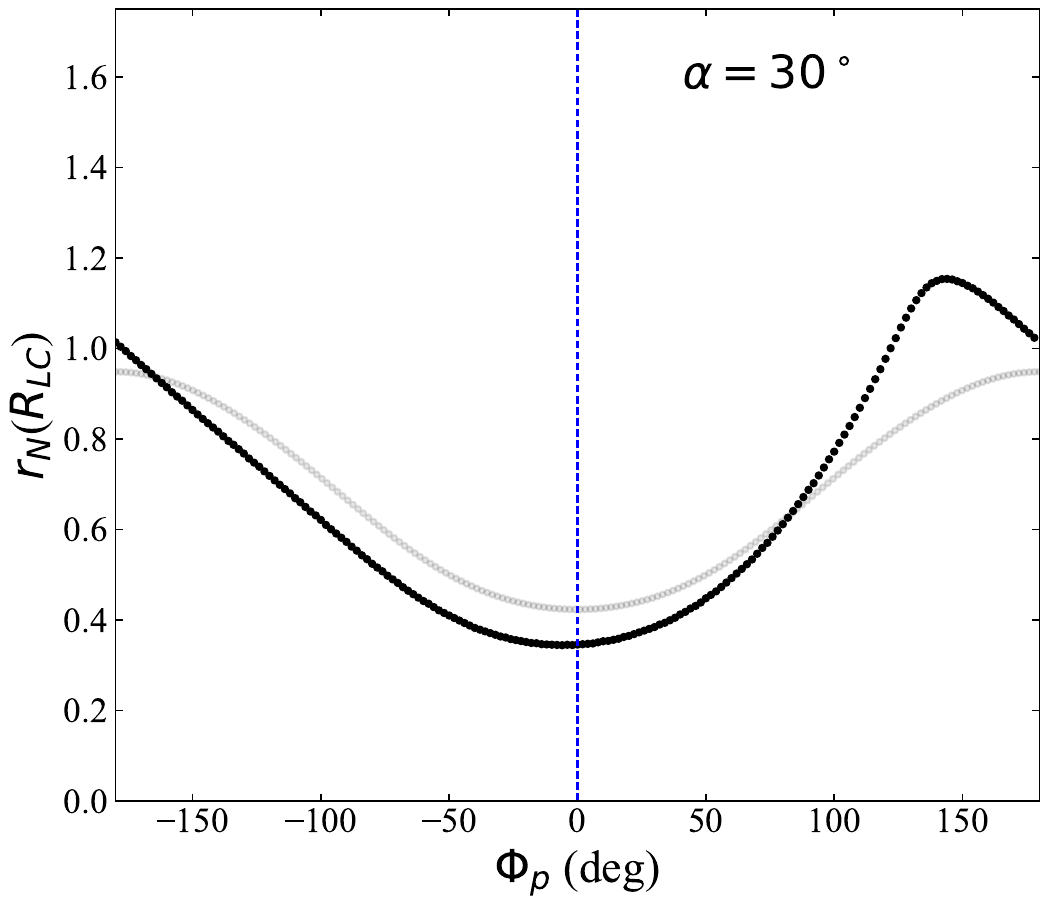} 
        \label{fig:sub1}
    \end{subfigure}
    \hfill 
    \begin{subfigure}[b]{0.245\textwidth} 
        \centering
        \includegraphics[width=\linewidth]{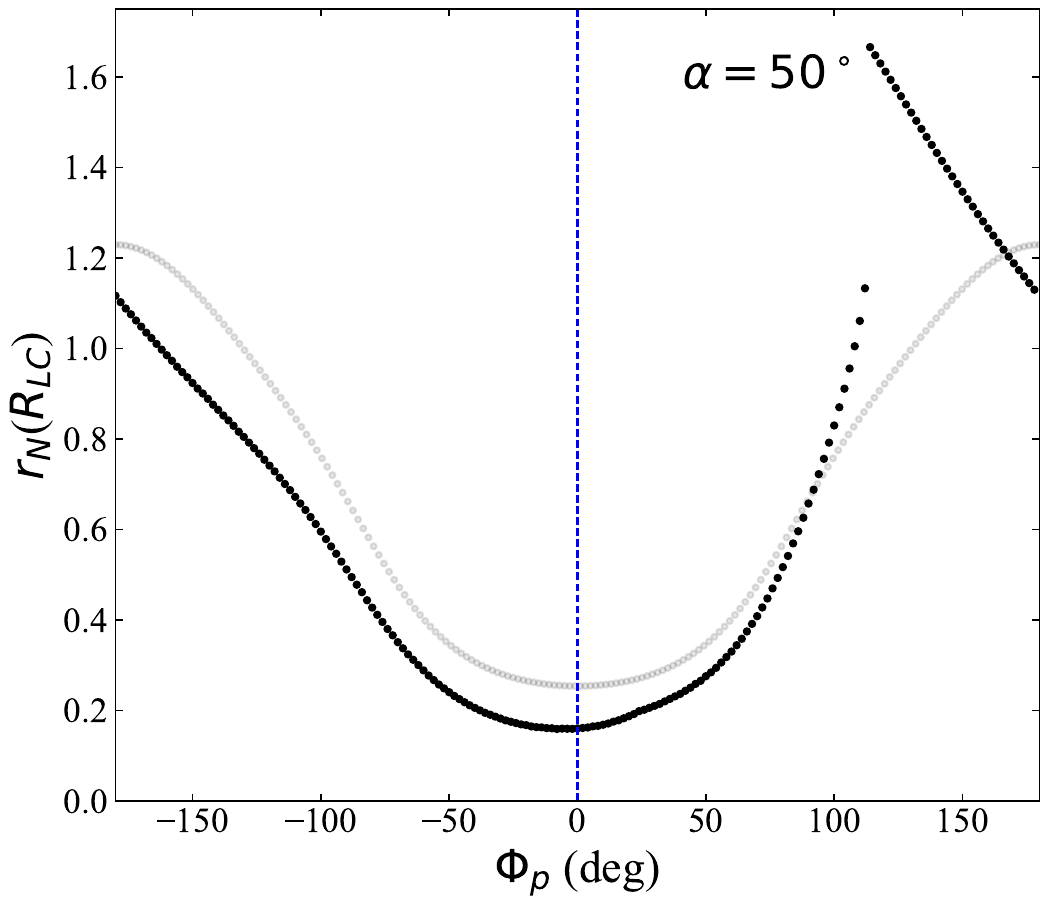} 
        \label{fig:sub1}
    \end{subfigure}
    \hfill
    \begin{subfigure}[b]{0.245\textwidth}
        \centering
        \includegraphics[width=\linewidth]{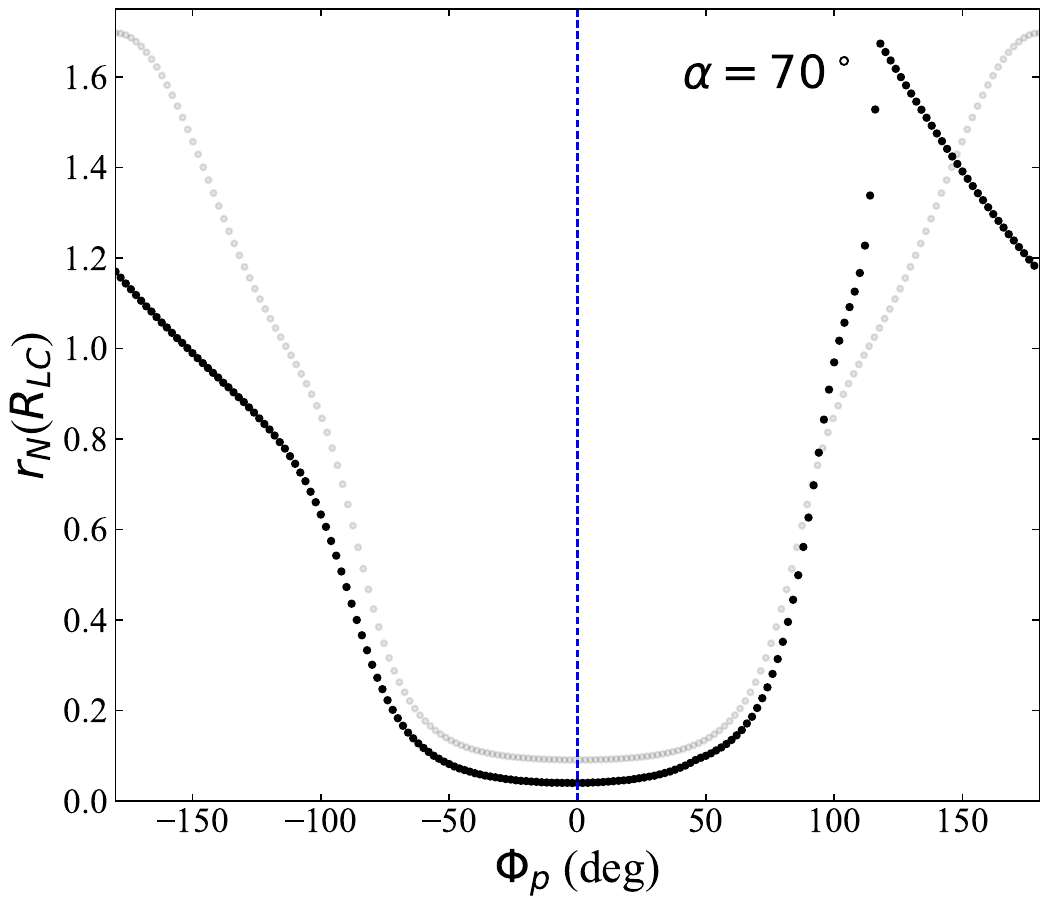}
        \label{fig:sub2}
    \end{subfigure}
    \caption{This figure shows the heights, $r_{\rm N}$, from the pulsar center to the intersection of the null charge surface with the last open field line, calculated using the Crab pulsar parameters and various $\alpha$, for both the static (light color) and rotating (dark color) dipole approximations. $\Phi_{\rm p}$ is the magnetic azimuth angle of footprints for these last open field lines. The blue vertical line marks where $\Phi_{\rm p} = 0^{\circ}$.}
    \label{fig:3}
\end{figure*}

The null charge surface is a fundamental structure in the pulsar magnetosphere. It plays an important role in several pulsar emission models, such as the outer gap model \citep{1986ApJ...300..500C,1986ApJ...300..522C} and the annular gap model \citep{2004ApJ...606L..49Q,2010MNRAS.406.2671D}.
The shape of this surface differs among magnetospheric approximations. Figure\, \ref{fig:3} compares the heights from the pulsar center to the intersection of the null charge surface with the last open field line for both the static and rotating dipole field approximations.
In the static dipole field approximation, this height is symmetric about the magnetic azimuth $\Phi_{\rm p} = 0^{\circ}$. In contrast, for the rotating dipole field approximation, it becomes progressively more asymmetric with respect to $\Phi_{\rm p} = 0^{\circ}$ as $\alpha$ increases. 
Furthermore, under the two approximations, there are also differences in the distribution range of heights for these heights.
This asymmetry and differences may make the specific position of the particle acceleration region and radiation height in models that depend on the geometry of the null charge surface change, such as the annular gap model \citep{2004ApJ...606L..49Q,2010MNRAS.406.2671D}.

\section{The annular gap model} \label{sec:3}
\subsection{Physics of model}
Critical field lines divide the open field line region of a pulsar into two parts: a core region (between the magnetic axis and the critical field lines) and an annular region (between the critical and last open field lines). If pulsars are bare quark stars or neutron stars with insufficient surface binding energy, both core‑gap and annular‑gap particle acceleration regions may develop\citep{2010MNRAS.406.2671D}. 
Within the static dipole field approximation, the annular polar region radius for an aligned rotator is given by $r_{\mathrm{ann}} = r_{\rm p} - r_{\mathrm{core}} = 0.26 R \left(\frac{\Omega R}{c}\right)$.
Thus, the annular acceleration region plays a significant role in pulsars with short spin periods\citep{2010MNRAS.406.2671D,2011ApJ...731....2D}, such as millisecond and young pulsars. 
In the annular gap model, high‑energy emission originates near the null charge surface, producing a fan‑beam gamma‑ray emission \citep{2010MNRAS.406.2671D,2011ApJ...731....2D,2007ChJAA...7..496Q}. Emission from both the core and annular regions can be simultaneously observed by the one observer if the inclination angle $\alpha$ and viewing angle $\zeta$ are suitable\citep{2004ApJ...616L.127Q}.

In the pulsar magnetosphere, charged particles near the light cylinder cannot corotate with the pulsar because of the light-speed limit and therefore must escape. As a result, the charge density near the light cylinder deviates from the GJ charge density, which will induce the generation of an accelerating electric field ($E_{||}$) parallel to the magnetic field. 
$E_{||}$ points in opposite directions in the annular and the core region, since these regions contain oppositely charged particles. Hence, $E_{||}$ vanishes at the boundary, i.e., the critical field lines between the annular and core regions, as well as along the last open field lines.
To replenish the escaped particles, the pulsar surface must supply fresh charged particles to the magnetosphere. These particles are accelerated in the annular and core regions and produce pulsed 
gamma-ray emission through a series of physical processes. This dynamic process is continually occurring in the magnetosphere.

\begin{figure*}[htbp]
    \centering
    \begin{subfigure}[b]{0.245\textwidth} 
        \centering
        \includegraphics[width=\linewidth]{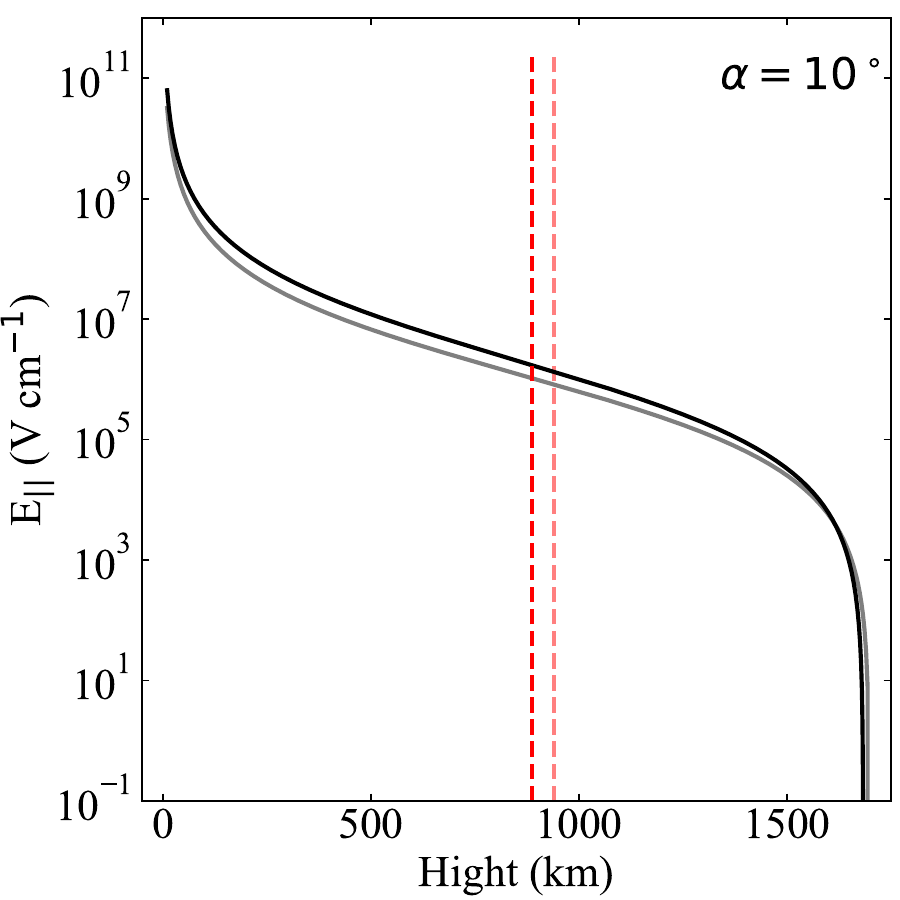} 
        \label{fig:sub1}
    \end{subfigure}
    \hfill
    \begin{subfigure}[b]{0.245\textwidth} 
        \centering
        \includegraphics[width=\linewidth]{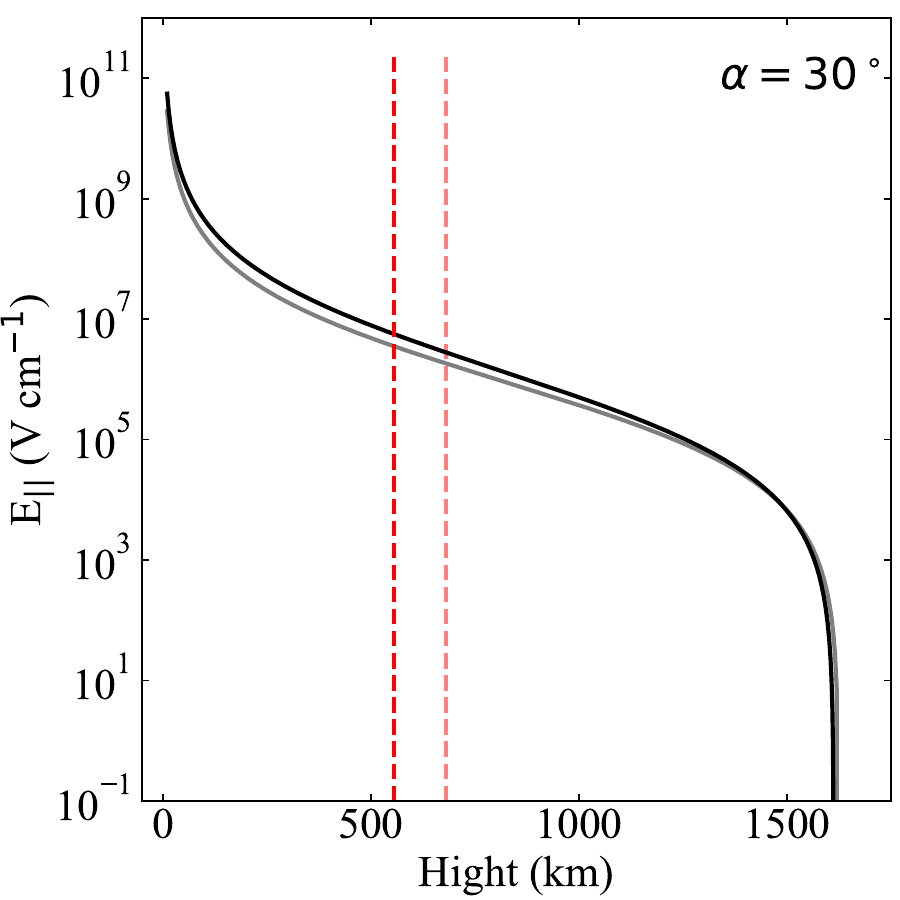} 
        \label{fig:sub1}
    \end{subfigure}
    \hfill 
    \begin{subfigure}[b]{0.245\textwidth} 
        \centering
        \includegraphics[width=\linewidth]{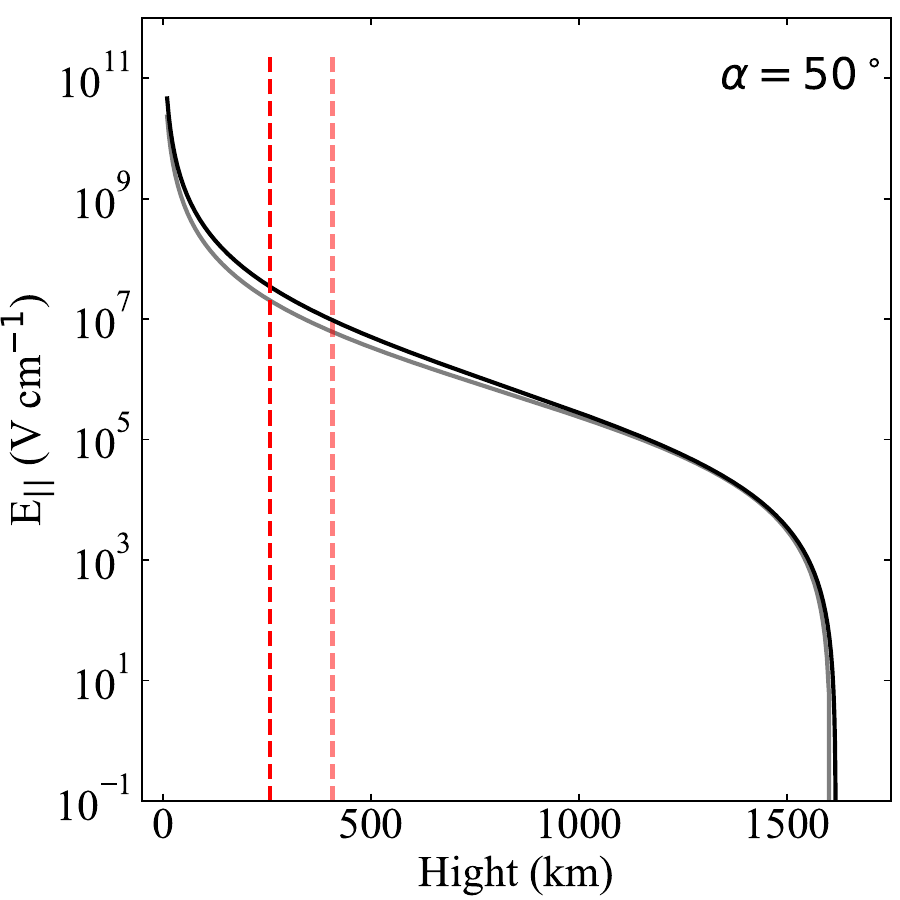} 
        \label{fig:sub1}
    \end{subfigure}
    \hfill
    \begin{subfigure}[b]{0.245\textwidth}
        \centering
        \includegraphics[width=\linewidth]{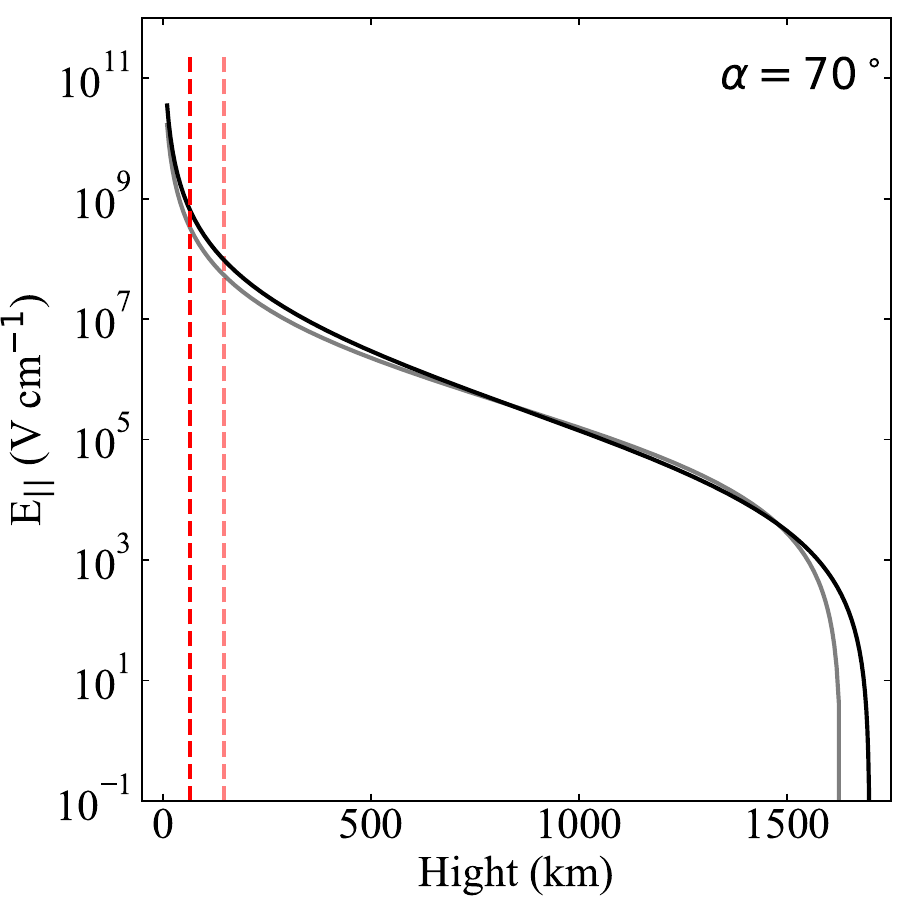}
        \label{fig:sub2}
    \end{subfigure}
    \caption{The figure shows the values of $E_{||}$ computed along a given field line using parameters of the Crab pulsar and different values of $\alpha$ under the rotating (dark shading) and the static (light shading) dipole field approximation. 
    The magnetic colatitude and magnetic azimuth of the footprint of this field line are $\frac{\Theta_{p,o}(0)+\Theta_{p,c}(0)}{2}$ and $0^\circ$, respectively, where  $\Theta_{p,o}(0)$ and $\Theta_{p,c}(0)$, respectively,  denote the magnetic colatitude of the footprints of the last open and the critical field line whose footprint magnetic azimuths are $0^\circ$, under the rotating dipole field approximation. 
    Two red vertical lines indicate the height from the pulsar center to the intersection of the null‑charge surface and the last open field line whose footprint has a magnetic azimuth of $0^\circ$ under the rotating (dark shading) and the static (light shading) dipole field approximation, respectively.}
    \label{fig:4}
\end{figure*}

The annular gap model posits that near the light cylinder, the charge density $\rho_{b}$ of outflowing particles, which move from the inner to the outer region and replenish lost charges, equals the local GJ charge density $\rho_{gj}$\citep{2010MNRAS.406.2671D}. That is, $\rho_b(S_{out}) = \rho_{gj}(S_{out})$, where $S_{out}$ is arc length calculated along a given magnetic field line from pulsar surface to light cylinder. Within a narrow magnetic flux tube, for any arc length $S < S_{out}$ along the same field line, the charge density satisfies $\rho_b(S) < \rho_{gj}(S)$. Therefore, an accelerating electric field, $E_{||}$, persists along the magnetic field line until $S$ reaches $S_{out}$. In the co-rotating frame, the derivative of $E_{||}$ with respect to $S$  is given by 
\begin{equation}
    \frac{dE_{||}}{dS} = 4\pi\left[\rho_b(S) - \rho_{gj}(S)\right],
\end{equation}
where $\rho_{gj}(S)\sim -\frac{\vec{\Omega}\cdot\vec{B}(S)}{2\pi c}$, and 
$\rho_{b}(S)$ can be derived from particle-number and magnetic-flux conservation in the flux tube:
\begin{equation}
    \rho_{b}(S) = \rho_{gj}(S_{out})\frac{B(S)}{B(S_{out})}
\end{equation}.
Using numerical methods, we calculated $E_{||}$ for both the static and rotating dipole field approximation, and these results are shown in Figure\,\ref{fig:4}. It can be seen that under these two approximations, the variation trend of $E_{||}$  with height is basically the same.

\subsection{Modeling of model}
Given that the trend of the $E_{||}$ with height is largely consistent under both the rotating and static dipole field approximations, following \cite{2011ApJ...731....2D}, we assume that the gamma‑ray emissivity $I(S)$ along one open field line in the annular and core regions follows a Gaussian distribution:
\begin{equation}
    I(S) = I_{peak}(\Theta_{\rm p},\Phi_{\rm p}) \text{exp} \left\{ -\frac{\left(S-S_{\mu}\right)^2}{2\sigma^2}\right\},
\end{equation}
where $\Theta_{\rm p}$ and $\Phi_{\rm p}$ are magnetic colatitude and magnetic azimuth angle for footprint of this open field line, respectively; $S$ and $S_\mu$ denote the arc lengths along this field line from the pulsar surface to an arbitrary emission point and to the point of peak emissivity, respectively; and $\sigma$ is a length scale for the emission region on each open field line in the annular region or the core region. 
% For the annular and core region, the heights, $r_{\rm p,ag}$ and $r_{\rm p,cg}$, of their peak emissivity points usually are different, and  are given by the equations
For the annulus and core regions, the heights corresponding to $S_\mu$, i.e., the heights of their peak emissivity points $r_{\rm p,ag}$ and $r_{\rm p,cg}$, are generally different and are given by the equations
\begin{equation}
    r_{\rm p,ag} = \lambda \kappa r_{\rm N}(\Phi_{\rm p}) + (1-\lambda) \kappa r_{\rm min}
\end{equation}
and 
\begin{equation}
    r_{\rm p,cg} = \epsilon r_{\rm p,ag}
\end{equation}
\citep{2011ApJ...731....2D}, respectively, where $r_N(\Phi_{\rm p})$ is the height from the pulsar center to the intersection point of the null charge surface and the last open field line whose footprint has magnetic azimuth of $\Phi_{\rm p}$; the model parameter $\kappa$ is the ratio of the peak emission height with $r_N(\Phi_{\rm p})$ \citep{2010MNRAS.406.2671D,2011ApJ...731....2D}; the model parameter $\lambda$ describe the deformation of emission location from a circle\citep{2006AdSpR..37.1988L}; $\epsilon$ also is a model parameter and represent emission height ratio for core and annular region. Since $r_N(\Phi_{\rm p})$ is no longer symmetric about $\Phi_{\rm p}=0^\circ$ under the rotating dipole field approximation, $r_{\text{min}} = \min[r_N(\Phi_{\rm p})]$, rather than $r_{\text{min}} = r_N(0)$ as in the static dipole field approximation.

For a set of open field lines whose footprints have same magnetic azimuth of $\Phi_{\rm p}$ in core or annular region, the peak emissivity $I_{peak}(\Theta_{\rm p},\Phi_{\rm p})$ are assumed to follow a Gaussian distribution in $\Theta_{\rm p}$, i.e.
\begin{equation}
    I_{peak}(\Theta_{\rm p},\Phi_{\rm p}) = I_0\exp \left\{-\frac{\left[\Theta_{\rm p}(\Phi_{\rm p})-\Theta_{\rm p}(a,\Phi_{\rm p})\right]^2}{2\sigma_{\theta}} \right\}
\end{equation}
where $I_0$ is a scaled emissivity; $\Theta_{\rm p}$ is used to label different open field lines in this set of open field lines; 
Here, for the core region, $\sin\left[\Theta_{\rm p}(a,\Phi_{\rm p})\right] = \sin\left[\Theta_{\rm p}(a_{\rm cg},\Phi_{\rm p})\right] = a_{\rm cg}\sin\left[\Theta_{p,c}(\Phi_{\rm p})\right]$, and for the annular region, $\sin\left[\Theta_{\rm p}(a,\Phi_{\rm p})\right] = \sin\left[\Theta_{\rm p}(a_{\rm ag},\Phi_{\rm p})\right] = \sin\left[\Theta_{p,c}(\Phi_{\rm p})\right] + a_{\rm ag}\left\{\sin\left[\Theta_{p,o}(\Phi_{\rm p})\right]-\sin\left[\Theta_{p,c}(\Phi_{\rm p})\right]\right\}$, where $\Theta_{p,c}(\Phi_{\rm p})$ and $\Theta_{p,o}(\Phi_{\rm p})$ are magnetic latitude of footprint for critical and last open field line in this set of open field lines, respectively; $a$ ($a_{\rm cg}$ or $a_{\rm ag}$) is a model parameter, $0<a<1$, and the value of $a$ may is not the same for different pulsars; $\sigma_{\theta}$ is transverse bunch scale for this set of open field lines. In addition, another model parameter $\Phi_{\rm p,cut}$ is introduced, which is a special magnetic azimuthal angle, and is used to constrain different regions for which Gaussian emissivities with different parameters are adopted \citep{2015ApJ...801..131D}.

To derive the ``photon sky-map" in the observer frame, we selected a set of open magnetic field lines in the core or/and annular regions, with their footprints uniformly distributed on the stellar surface. These footprints were spaced at intervals of $1^{\circ}$ in magnetic azimuth $\Phi$ and 0.0001 radians in magnetic colatitude $\Theta$.
Then we calculated the emission intensity $I(r, \Theta, \Phi)$ and direction $\vec{n}_B(r, \Theta, \Phi)$ for emission points on those open field lines in the corotation frame. We used equation (1) from \cite{2003ApJ...598.1201D} to consider the aberration effect and transform the emission direction $\vec{n}_B(r, \Theta, \Phi)$ to  $\vec{n}_v(r, \theta_{em}, \phi_{em})$ in the inertial observer frame. The phase shift caused by light travel time delays was also taken into account and computed by the equation $\delta \phi = \vec{r}\cdot\vec{n}_v/R_{\rm LC}$ \citep{2003ApJ...598.1201D}. Combining  $\phi = -\phi_{em}-\delta \phi$ and $\zeta = \arccos\left\{n_{v,z}/\sqrt{n_{v,x}^2+n_{v,y}^2+n_{v,z}^2}\right\}$, the ``photon sky-map" $I(\phi,\zeta)$ can be made.

\begin{table*}
\centering
\caption{Best-fit parameters for the gamma-ray ($>0.1$ GeV) light curves of the sample pulsars.}\label{tab:parameters}
\setlength{\tabcolsep}{3.7pt}
\begin{tabular}{@{}ccccccccc@{\quad}cccccccc@{}}
\toprule
\multirow{5}{*}{PSR} & \multicolumn{8}{c}{Global and Geometric Parameters} & \multicolumn{8}{c}{Emission Region Scale Parameters}\\
\cmidrule(lr){2-9} \cmidrule(lr){10-17} &  &  & & & & & &  &
 \multicolumn{2}{c}{$\sigma_{\rm A}$} & \multicolumn{2}{c}{$\sigma_{\rm \theta,A}$} &
\multicolumn{2}{c}{$\sigma_{\rm C}$} & \multicolumn{2}{c}{$\sigma_{\rm \theta,C}$} \\
\cmidrule(lr){2-9} \cmidrule(lr){10-11} \cmidrule(lr){12-13} \cmidrule(lr){14-15} \cmidrule(lr){16-17} 
 & $\alpha$ & $\zeta$ & $a_{\rm ag}$ & $a_{\rm cg}$ & $\kappa$ & $\lambda$ & $\epsilon$ & $\Phi_{\rm cut}$ &
$\sigma_{\rm A1}$ & $\sigma_{\rm A2}$ & $\sigma_{\rm \theta,A1}$ & $\sigma_{\rm \theta,A2}$ &
$\sigma_{\rm C1}$ & $\sigma_{\rm C2}$ & $\sigma_{\rm \theta,C1}$ & $\sigma_{\rm \theta,C2}$ \\
 & (deg) & (deg) & & & & & & (deg) &
($R_{\rm LC}$) & ($R_{\rm LC}$) & ($rad$) & ($rad$) &
($R_{\rm LC}$) & ($R_{\rm LC}$) & ($rad$) & ($rad$)\\

\midrule
J0631$+$1036 & 16 & 18 & 0.05 & 0.88 & 0.60 & 0.00 & 1.30 & 0 &
0.030 & 0.030 & 0.0040 & 0.0040 & 0.010 & 0.010 & 0.0015 & 0.0010 \\
J1709$-$4429 & 148 & 136 & 0.50 & -- & 0.85 & 0.94 & -- & 15 &
0.015 & 0.025 & 0.0013 & 0.0013 & -- & -- & -- & -- \\
J1048$-$5832 & 155 & 150 & 0.01 & 0.87 & 0.60 & 0.90 & 1.20 & 0 &
0.050 & 0.050 & 0.0019 & 0.0019 & 0.020 & 0.020 & 0.0010 & 0.0010 \\
\bottomrule
\label{table_1}
\end{tabular}
\vspace{1mm}
\begin{minipage}{\textwidth}
\small  
\textbf{Note:}$\sigma_{\rm A1}$ and $\sigma_{\rm A2}$ are the length scales of the radiation regions on open field lines in the annular region, corresponding to footprint magnetic azimuthal angles $\Phi_{\rm p}$ in the ranges $-180^\circ < \Phi_{\rm p} < \Phi_{\rm cut}$ and $\Phi_{\rm cut} < \Phi_{\rm p} < 180^\circ$, respectively. Similarly, $\sigma_{\rm C1}$ and $\sigma_{\rm C2}$ are the corresponding length scales in the core region for the same $\Phi_{\rm p}$ ranges.
$\sigma_{\rm \theta,A1}$ and $\sigma_{\rm \theta,A2}$ are the transverse bunch scales for field lines in the annular region over the same azimuthal ranges, while $\sigma_{\rm \theta,C1}$ and $\sigma_{\rm \theta,C2}$ are the transverse bunch scales for the core region, also defined over the corresponding $\Phi_{\rm p}$ intervals.
\end{minipage}
\end{table*}

\section{result} \label{sec:4}
In this section, we shall apply the annular gap model to the observations. Three young pulsars(PSRs J0631$+$1036, J1709$–$4429, and J1048$–$5832) that show different numbers of peaks in their gamma-ray light curves above 0.1 GeV are selected from 3PC. Young pulsars generally exhibit strong polarization, and their polarization position angle(PPA) traverses are often consistent with the rotating vector model\citep[RVM, ][]{1969ApL.....3..225R,2001ApJ...553..341E}. This agreement allows us to constrain their viewing geometries($\alpha$ and $\zeta$). 
We obtained the polarization data at 1.4 GHz of these three pulsars from \cite{2018MNRAS.474.4629J}, and used these to constrain their viewing geometries.  
Based on the constrained viewing geometries, we simulated these three pulsars’ gamma-ray light curves above 0.1 GeV using the annular model. The specific details and results are as follows.

\subsection{Constraining viewing geometry} \label{sec:4.1}
Radio polarization observations are particularly useful for constraining the viewing geometries of pulsars. The radio emission may originate from three distinct regions: the traditional low-altitude polar cap region \citep{1975ApJ...196...51R}, the high-altitude outer magnetosphere near the light cylinder \citep{2005Ap&SS.297..101M}, or the current sheet region beyond the light cylinder \citep{2025arXiv251005778K}.
Pulsars with radio emission originating from the polar cap region typically exhibit an S-shaped PPA traverse in their radio profiles. The three selected pulsars all display this characteristic. The S-shaped PPA traverse can be interpreted within the framework of the RVM. This model states that the PPA traverse corresponds to the changing orientation of the planes containing the diverging dipolar magnetic field lines as the pulsar’s beam sweeps across our line of sight.

According to the RVM, the PPA, $\psi$, is defined as a function of the pulse longitude and given by the equation
\begin{equation}
    \tan \big ( \psi - \psi_0 \big ) = \frac{\sin \alpha \sin \big (\phi - \phi_0 \big )}{\sin \zeta \cos \alpha - \cos \zeta \sin \alpha \cos (\phi -\phi_0)},
\end{equation}
where $\alpha$ is the inclination angle between the magnetic axis and the rotation axis; $\zeta = \alpha + \beta$ is the angle between the line-of-sight and the rotation axis, $\beta$ is the impact parameter, namely, the angle between the line-of-sight and the magnetic axis at the position of closest approach; $\psi_0$ and $\phi_0$ are the PPA and pulse longitude, respectively, of the steepest point of this PPA traverse. In this formulation, $\psi$ increases clockwise in the sky, which is opposite to the astronomical convention \citep[PSR/IEEE convention,][]{2010PASA...27..104V}.  To align with that convention, we multiply the numerator in this formulation by $-1$ when this formulation is used.

In principle, fitting the PPA traverse to the RVM model can determine viewing geometries ($\alpha$ and $\zeta$). However, in practice, the viewing geometries derived by this method usually carry significant uncertainty because $\alpha$ and $\zeta$ are covariant, and the proportion of the radio emission window in the pulse period is limited \citep{2004A&A...421..215M}. Thus, the viewing geometries is generally only constrained rather than precisely determined.
We fitted the PPA traverses of our pulsar sample using the RVM and computed the reduced chi-square ($\chi^2$) values for the fits over a range of $\alpha$ and $\zeta = \alpha + \beta$ combinations, thereby placing constraints on the viewing geometries.

\subsection{Modeling result}
We have simulated gamma-ray light curves above 0.1 GeV for our pulsar sample using the annular gap model and calculated the corresponding radiation heights. The simulation procedure is as follows. First, the viewing geometries were constrained using archival polarization data, and the phase of the steepest point of the PPA traverse(i.e., the phase of the fiducial plane) was determined. Then, based on the time-aligned radio light curves provided by the 3PC, we identified the position of the fiducial plane within the gamma-ray light curves. A combination of ($\alpha$, $\zeta$) was selected from within a $3\sigma$ confidence region, and other model parameters were adjusted accordingly. This process was repeated iteratively until the time-aligned gamma-ray light curve above 0.1 GeV was simulated. 

\begin{figure*}[htbp]
    \centering
    \begin{subfigure}[b]{0.475\textwidth} 
        \centering
        \includegraphics[width=\linewidth]{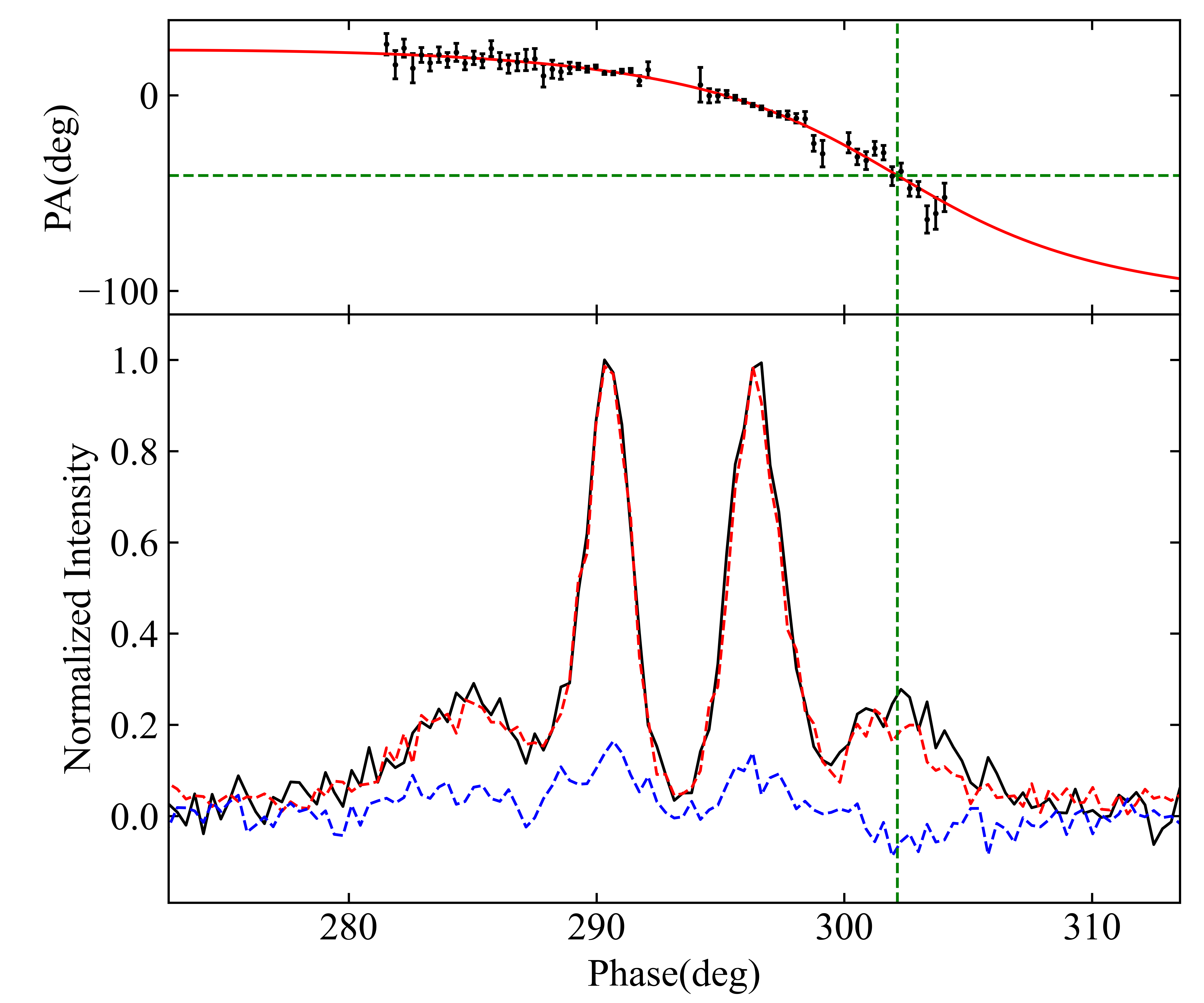} 
    \end{subfigure}
    \hfill
    \begin{subfigure}[b]{0.495\textwidth} 
        \centering
        \includegraphics[width=\linewidth]{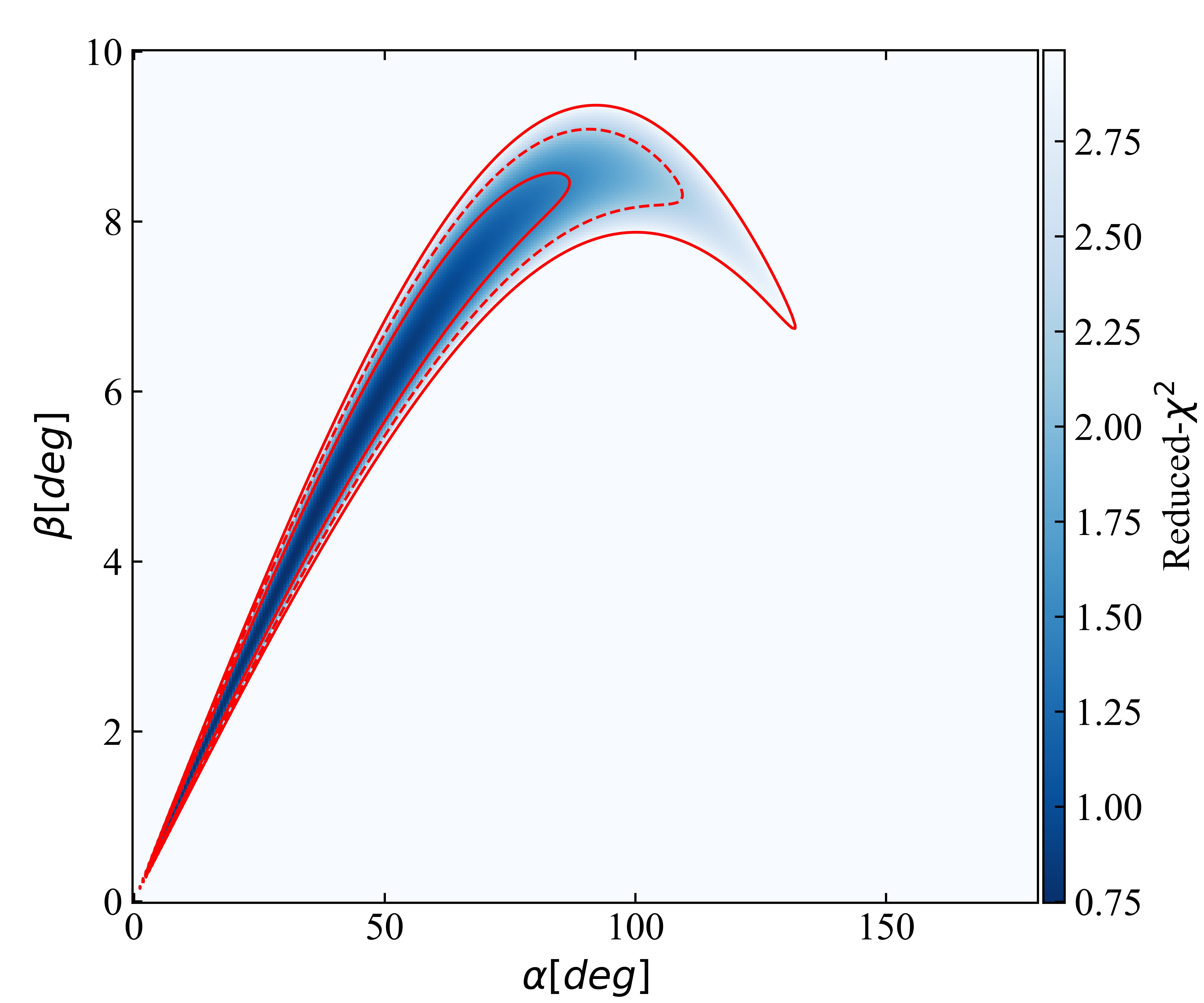} 
    \end{subfigure}
    \hfill 
    \begin{subfigure}[b]{0.495\textwidth} 
        \centering
        \includegraphics[width=\linewidth]{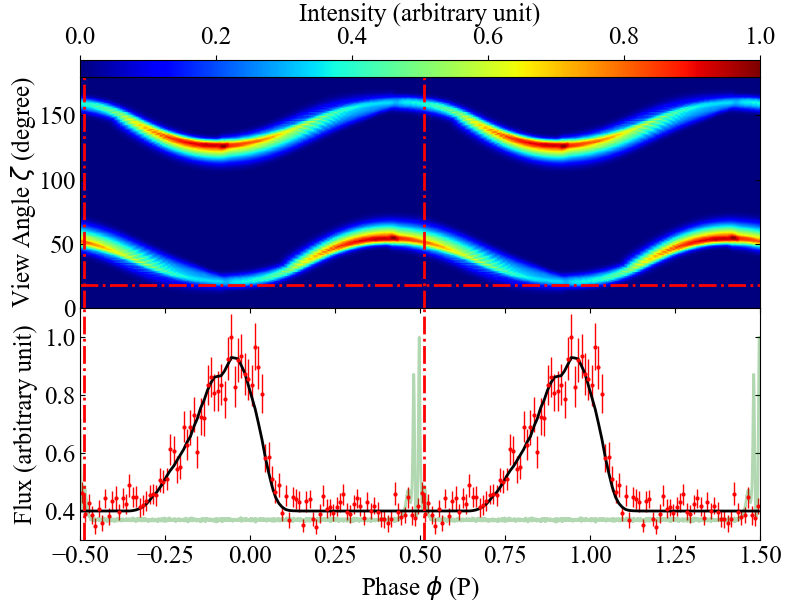} 
    \end{subfigure}
    \hfill
    \begin{subfigure}[b]{0.49\textwidth}
        \centering
        \includegraphics[width=\linewidth]{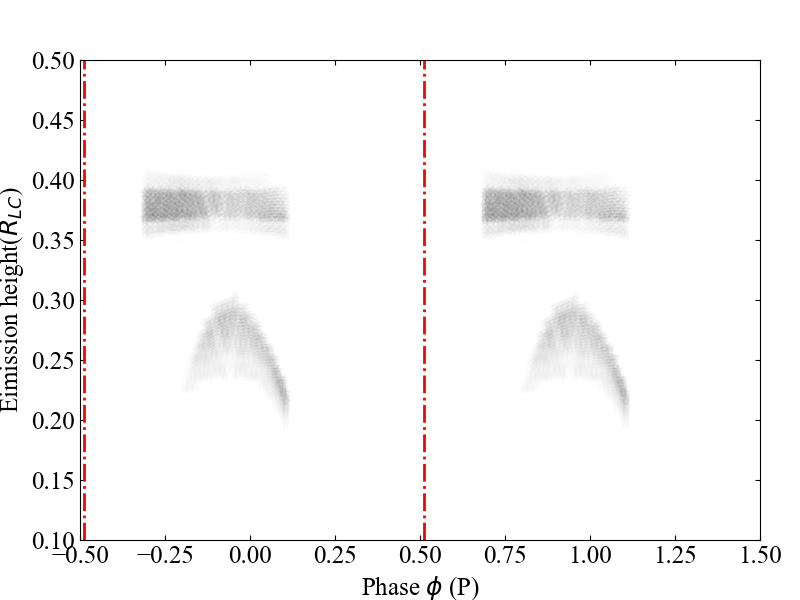}
    \end{subfigure}
    \caption{ Joint fitting of radio polarization and gamma-ray light curves for PSR J0631+1036. Top-left: The lower panel displays the total intensity (black solid line), total linear polarization (red dashed line), and circular polarization (blue dashed line). The upper panel shows the PPA as black dots with error bars, along with the best-fit RVM result plotted as a red curve. Top-right: The reduced $\chi^2$ of the fit is represented by the blue color scale, with red contours marking the 1$\sigma$, 2$\sigma$, and 3$\sigma$ confidence levels. Bottom-left: The upper panel presents a gamma-ray photon sky map spanning two rotational periods, and the lower panel shows the observed (red dots with error bars) and simulated(black solid line) gamma-ray light curves as well as the time-aligned radio light curves(light-green solid line), spanning two rotational periods. The vertical and horizontal red dashed lines represent the phase of the fiducial plane and the line-of-sight angle, respectively. Bottom-right panel: the gamma ray radiation height corresponding to the simulation is shown.}
    \label{fig:J0631+1036}
\end{figure*}

\textbf{PSR J0631$+$1036}. This pulsar has a period of 288 ms\citep{2024MNRAS.530.1581K} and a radio/gamma-ray phase lag($\delta$) of $\sim 0.51P$ \citep{2023ApJ...958..191S}. Its radio pulse profile and corresponding PPA (top-left panel), the $\chi^2$ of the radio polarization fitting (top-right panel), and the gamma-ray light curve(bottom-left panel) are shown in Figure~\ref{fig:J0631+1036}. The gamma-ray light curves exhibit a single-peaked profile, whereas the radio light curves show four distinct peaks. We simulated the gamma-ray light curves using annular and core regions, and the results are presented in the bottom-left panel of Figure~\ref{fig:J0631+1036}; the simulation reproduces the observed gamma-ray light curve well. Thus, the single-peak feature in the gamma-ray light curve may be attributed to the line of sight traversing the overlapping edge of the gamma-ray radiation beams from the core and annular region. The relevant parameters used in the simulation are listed in Table~\ref{table_1}. As shown in the bottom-right panel of Figure\,\ref{fig:J0631+1036}, the corresponding radiation heights are concentrated mainly between $0.2-0.3R_{\rm LC}$ and $0.35-0.4R_{\rm LC}$. 

\begin{figure*}[htbp]
    \centering
    \begin{subfigure}[b]{0.475\textwidth} 
        \centering
        \includegraphics[width=\linewidth]{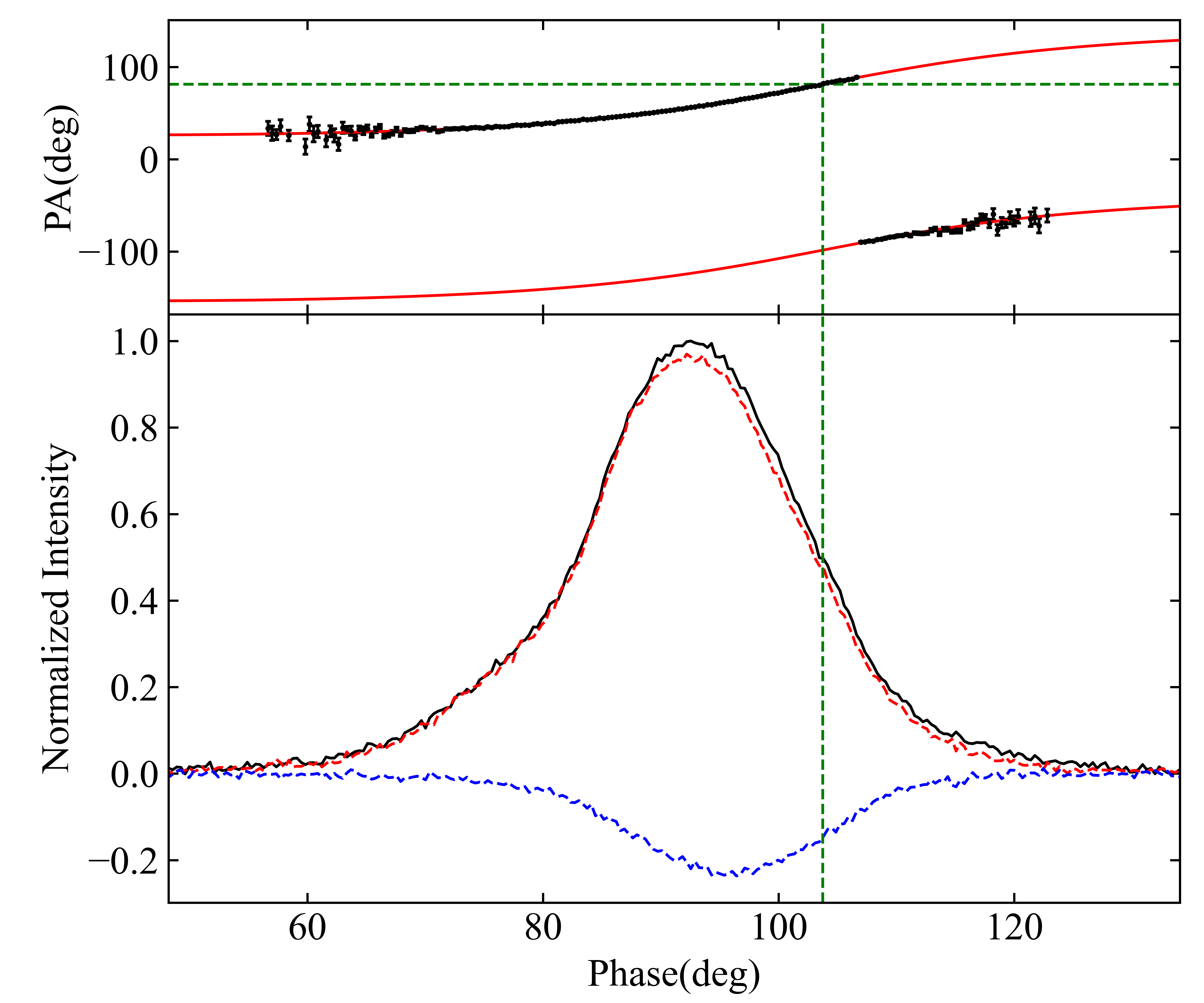} 
    \end{subfigure}
    \hfill
    \begin{subfigure}[b]{0.49\textwidth} 
        \centering
        \includegraphics[width=\linewidth]{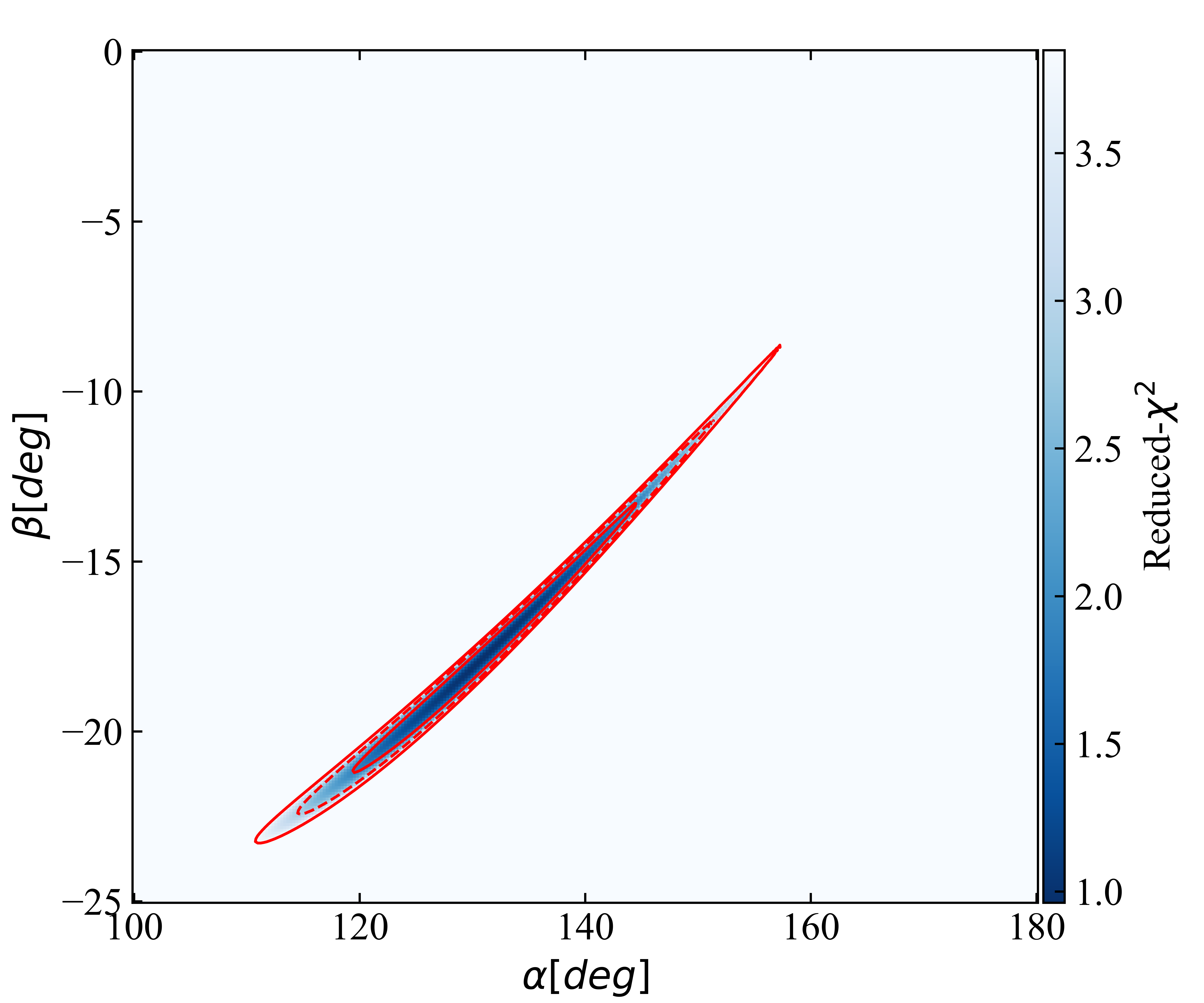} 
    \end{subfigure}
    \hfill 
    \begin{subfigure}[b]{0.495\textwidth} 
        \centering
        \includegraphics[width=\linewidth]{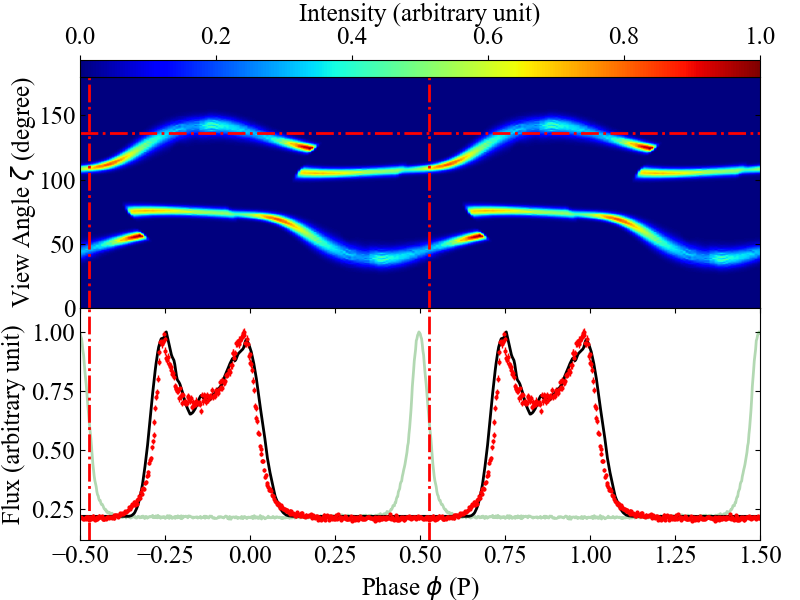} 
    \end{subfigure}
    \hfill
    \begin{subfigure}[b]{0.495\textwidth}
        \centering
        \includegraphics[width=\linewidth]{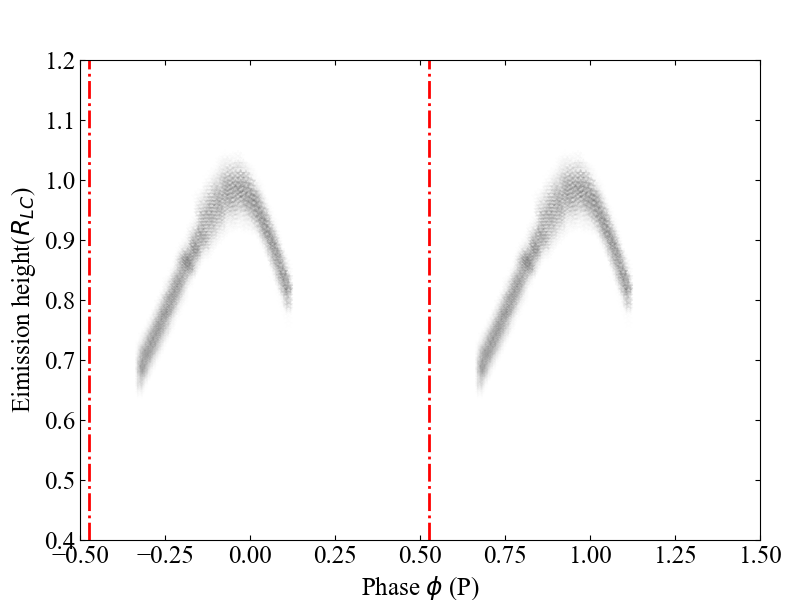}
    \end{subfigure}
    \caption{Same as Figure\,\ref{fig:J0631+1036}, but for PSR J1709$-$4429}
    \label{fig:J1709-4429}
\end{figure*}

\textbf{PSR J1709$-$4429}. This pulsar was discovered in a high-frequency survey of the southern Galactic plane \citep{1992MNRAS.255..401J}, with a period of $\sim$103 ms \citep{2024MNRAS.530.1581K} and a $\delta$ value of $\sim 0.236P$ \citep{2023ApJ...958..191S}. Its viewing geometries are better constrained than those of the other pulsars in our sample (top-right panel of Figure \ref{fig:J1709-4429}). The radio pulse profile shows a single-peaked structure (top-left panel of Figure \ref{fig:J1709-4429}), while the gamma-ray light curve exhibits a double-peaked structure (bottom-left panel of Figure \ref{fig:J1709-4429}).  This pulsar has a relatively large impact angle $\beta$ (top-right panel of Figure \ref{fig:J1709-4429}), and we likely do not observe any emission from the core region. Thus, we simulate its gamma-ray light curve using only the annular region, and the simulation result is presented in the bottom-left panel of Figure \ref{fig:J1709-4429}.  The main features of the observed gamma-ray light curve are reproduced well. The double-peaked feature in its gamma-ray light curve may be caused by the line of sight passing through the gamma-ray beam from the annular region. The viewing geometries and model parameters adopted in the simulation are listed in Table \ref{table_1}. Simulation results show that the gamma-ray emission originates primarily from heights between $0.65 R_{\rm LC}$ and $1.05 R_{\rm LC}$, as shown in the bottom-right Figure\,\ref{fig:J1709-4429}. 

\begin{figure*}[htbp]
    \centering
    \begin{subfigure}[b]{0.475\textwidth} 
        \centering
        \includegraphics[width=\linewidth]{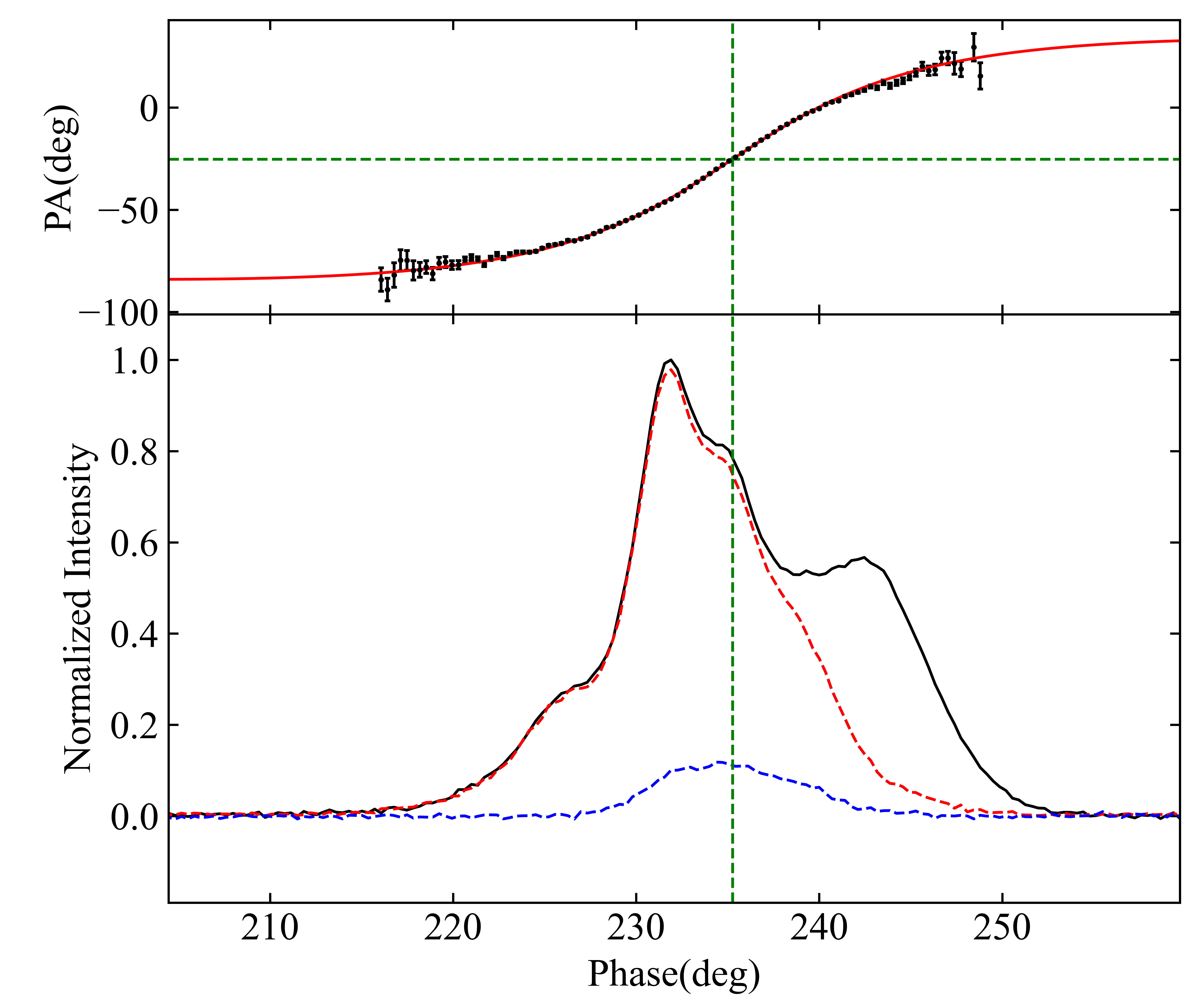} 
    \end{subfigure}
    \hfill
    \begin{subfigure}[b]{0.49\textwidth} 
        \centering
        \includegraphics[width=\linewidth]{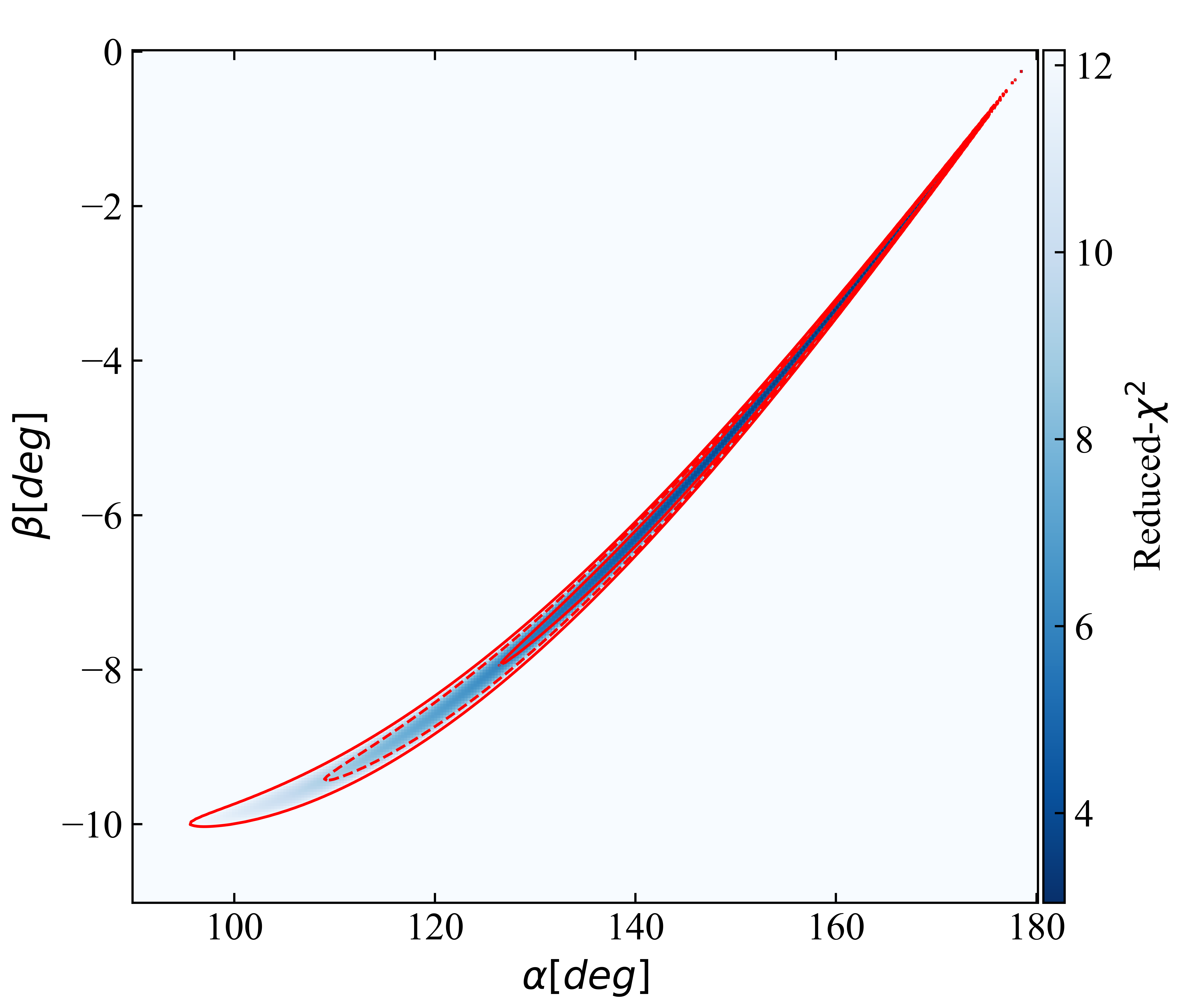} 
    \end{subfigure}
    \hfill 
    \begin{subfigure}[b]{0.495\textwidth} 
        \centering
        \includegraphics[width=\linewidth]{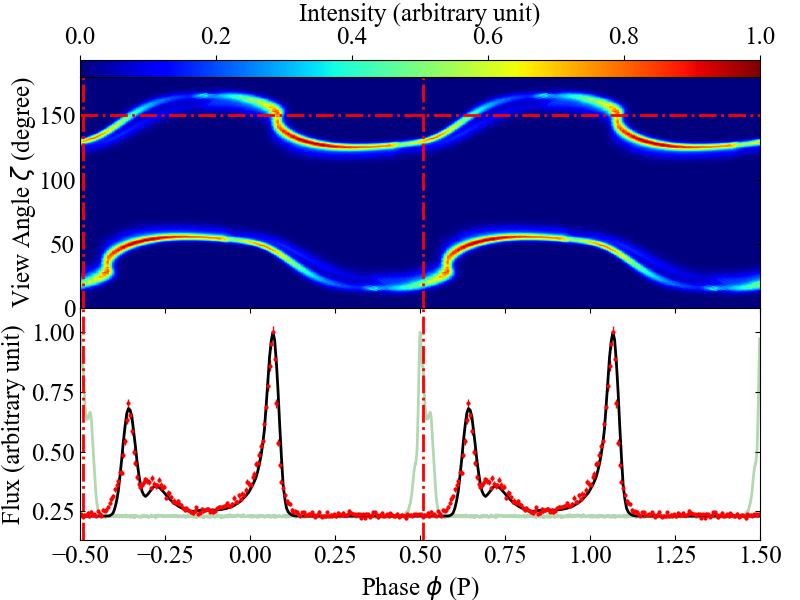} 
    \end{subfigure}
    \hfill
    \begin{subfigure}[b]{0.495\textwidth}
        \centering
        \includegraphics[width=\linewidth]{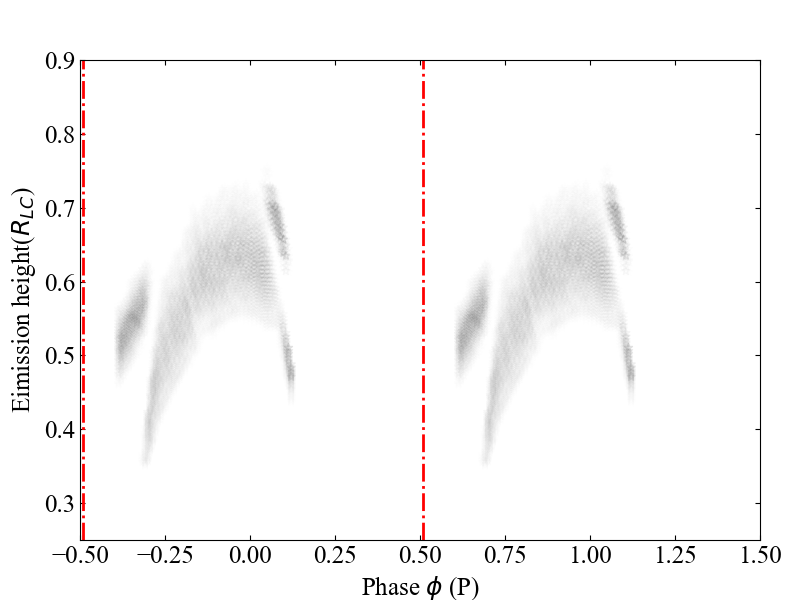}
    \end{subfigure}
    \caption{Same as Figure\,\ref{fig:J0631+1036}, but for PSR J1048$-$5832}
    \label{fig:J1048-5832}
\end{figure*}

\textbf{PSR J1048$-$5832}. It is a bright gamma-ray pulsar with a spin period of 123.7 ms\citep{2024MNRAS.530.1581K}, $\delta$ of $\sim 0.147P$\citep{2023ApJ...958..191S}, and a characteristic age of 20.4 thousand years.
As shown in Figure\,\ref{fig:J1048-5832}, its radio pulse profile displays four peaks, while the gamma‑ray light curve shows three peaks. 
This requires that, in simulating the gamma‑ray light curve, we must use both the annular and core regions. 
Results of our simulations are shown in the bottom-left panel of Figure\,\ref{fig:J1048-5832}, and show that the main observed features of the light curve are reproduced well.
Therefore, we propose that the observed gamma‑ray emission originates from contributions of both the annular and core regions, and that their gamma‑ray beams partially overlap. The line of sight passing through this overlapped beam produces a three‑peaked gamma‑ray light curve. The parameters adopted in the simulation are listed in Table\,\ref{table_1}. Within this model, the gamma‑ray emission of PSR J1048$-$5832 originates primarily from heights between $0.35R_{\rm LC}$ and $0.75R_{\rm LC}$(see in the bottom-right panel of Figure\,\ref{fig:J1048-5832}).

\section{Discussions and Conclusions} \label{sec:5}
If the radial distance is not too far from the pulsar surface, the pulsar’s magnetic field can be approximated as a static dipolar structure\citep{2005ApJ...630..454M}. Since the gamma-ray emission region in the annular gap model does not lie at a radial distance far from the pulsar surface, \citet{2010MNRAS.406.2671D} argued that the annular gap model exhibits weak dependency on the magnetospheric approximation and thus adopted a static dipole field as an approximation for the pulsar magnetosphere. 
However, as the definition of the annular gap model relies on the last open field lines, the critical field lines, and the null charge surface\citep{2004ApJ...606L..49Q}, which are directly or indirectly dependent on the magnetic field in the vicinity of the light cylinder, the annular gap model may have a non-negligible dependency on the magnetospheric approximation, especially when the magnetic inclination $\alpha$ is large. We employed a rotating dipole field as an approximation for the magnetic field in the pulsar magnetosphere.
Compared with the static dipole field approximation, the pulsar’s open field region increases in size, and the last open field lines, critical field lines, and null charge surface no longer remain symmetric with respect to the fiducial plane. In the context of the annular gap model, this leads to the annular region, the core region, and their peak emissivity region not being symmetric with respect to the fiducial plane, thereby producing a pulsar radiation beam that is asymmetric even after removing aberration and time delay effects.

If observed radiation from pulsars originates solely from a single magnetic pole, our model can generate light curves with at most four peaks.  Thus, it cannot explain light curves that exhibit more than four peaks.  Pulsars whose gamma-ray light curves show more than four peaks are exceedingly rare; only three are known (PSRs J0030$+$0451, J1536$-$4948, and J2241$-$5236)\citep{2023ApJ...958..191S}.  Among these three, the gamma-ray light curves of two pulsars are interrupted by off-pulse regions, resulting in discontinuous gamma emission components, and the isolated gamma emission components are markedly asymmetric.  
Therefore, we speculate that the gamma-ray emission from such pulsars may also originate not only from the region described by the annular gap model, but also potentially from other regions. 
\cite{2025arXiv251005778K} speculated a similar conclusion, that the radio radiation originates from multiple radiation regions, in their research on millisecond pulsars.

During the process of using the annular model to simulate the gamma-ray light curves above 0.1 GeV of pulsars J0631$+$1036, J1709$-$4429, and J1048$-$5832, we found that employing several other sets of slightly varied geometrical and model parameters could approximately reproduce the main characteristics of these pulsars’ gamma-ray light curves. For the sake of brevity, these details are not listed individually in the text. If one wants to more accurately determine the viewing geometries and model parameters of these pulsars, it may be necessary to integrate radiative features from additional wavelength bands for comprehensive constraints. In subsequent studies, we will incorporate data from more wavelength bands and conduct further in-depth investigations.

In recent years, advances in numerical techniques and increased computational power have enabled extensive studies and simulations of the pulsar magnetosphere, leading to the development of several pulsar magnetospheric models. These include force-free magnetospheres\citep[e.g.][]{2016MNRAS.455.3779P}, resistive magnetospheres \citep[e.g.][]{2017ApJ...850..205K}, and kinetic particle-in-cell (PIC) magnetospheres\citep[e.g.][]{2015ApJ...801L..19P}. 
The kinetic PIC magnetosphere represents a more realistic, first-principles description of the pulsar magnetosphere, revealing more complex magnetospheric structures and plasma physical processes. In light of these advances, we may revisit and refine the annular gap model by incorporating more realistic pulsar magnetosphere models and plasma physical processes in future studies.

By using a rotating dipole field that is more realistic than a static dipole field as an approximation of the magnetic field in the pulsar magnetosphere, we revisit the annular gap model and apply it to observations. 
Under this approximation, the open region, including the core region and the annular region, has become larger(compared with the static dipole approximation), and the core region, the annular region, and their peak emissivity region are not symmetrical about the fiducial plane, especially for large values of $\alpha$. We selected three young pulsars that have distinct gamma-ray light-curve profiles: PSRs J0631$+$1036 (single-peaked), J1709$-$4429(double-peaked), and J1048$-$5832 (three-peaked).
By analyzing radio polarization data from these pulsars, we constrained their viewing geometries. 
With these constraints and the annular gap model under the rotating dipole field approximation, we successfully reproduce the key morphological features of the gamma-ray light curves above 0.1 GeV for these three pulsars. 
Our simulations show that the primary gamma-ray emission region for PSR J0631$+$1036 may lies at distances of $0.2-0.3 R_{\rm LC}$ and $0.35-0.4R_{\rm LC}$ from the stellar center, for PSR J1709$-$4429 it may is located at $0.65–1.05R_{\rm LC}$, and for PSR J1048$-$5832 may at $0.35–0.75R_{\rm LC}$. This model provides a method for interpreting pulsar high-energy emission, advancing our understanding of this radiation.
\section*{Acknowledgement} 
This work is supported by the National Natural Science Foundation of China (Nos. 12273008), the National SKA Program of China (Nos.2022SKA0130100, 2022SKA0130104), the Natural Science and Technology Foundation of Guizhou Province (No. [2023]024), and the Guizhou Provincial Basic Research Program (Natural Science) (No. Qiankehejichu-MS(2025)263).

\bibliography{sample701}{}
\bibliographystyle{aasjournalv7}
\end{document}